\documentclass[11pt]{article}
\makeatletter
\def\@cite#1#2{$^{\hbox{\scriptsize{#1\if@tempswa , #2\fi})}}$}
\def\thebibliography#1{
 \section*{References\@mkboth{References}{References}}
 \list{\arabic{enumi})}
   {\settowidth{\labelwidth}{[#1]}
    \leftmargin=\labelwidth
    \advance \leftmargin by \labelsep
    \usecounter{enumi}}
 \def\newblock{\hskip .11em plus .33em minus .07em}
 \sloppy
 \clubpenalty=4000 \widowpenalty=4000
 \sfcode`\.=1000\relax}

\makeatother
\usepackage{amsfonts}
\usepackage{amssymb}
\usepackage{amsmath}
\usepackage[dvips]{graphicx}
\usepackage{psfrag}
\renewcommand{\theequation}{\Roman{section}.\arabic{equation}}
\renewcommand{\thefootnote}{\fnsymbol{footnote}}
\newcommand{\scsc}{\scriptscriptstyle}
\newcommand{\dps}{{\cal D} \psi}

\newcommand{\dmd}{\partial_{\mu}}

\newcommand{\beq}{\begin{eqnarray}}
\newcommand{\eeq}{\end{eqnarray}}

\begin{document}
%%%%%%%%%%%%%%%%%%%%%%%%%%%%%%%%%%%%%%%%%%%%%%%%%%%%%%%%%%%%%%%%%%%%%%%%%
%%%%%%%%%%%%%%%%%%%% TITLE PAGE %%%%%%%%%%%%%%%%%%%%%%%%%%%%%%%%%%%%%%%%%
%%%%%%%%%%%%%%%%%%%%%%%%%%%%%%%%%%%%%%%%%%%%%%%%%%%%%%%%%%%%%%%%%%%%%%%%%
\begin{titlepage}
%%%%%%%%%%%%%%%%%%%%%%%%%%%%%%%%%%%%%%%%%%%%%%%%%%%%%%%%%%%%%%%%%%%%%%%%%
%%%%%%%%%%%%%%%%%%%%%%%%%%%%%%%%%%%%%%%%%%%%%%%%%%%%%%%%%%%%%%%%%%%%%%%%%
\begin{flushright}
HUPD-0110\\
hep-th/0107033
\end{flushright}
%%%%%%%%%%%%%%%%%%%%%%%%%%%%%%%%%%%%%%%%%%%%%%%%%%%%%%%%%%%%%%%%%%%%%%%%%
\begin{center}
\vspace*{1.6cm}
%%%%%%%%%%%%%%%%%%%%%%%%%%%%%%%%%%%%%%%%%%%%%%%%%%%%%%%%%%%%%%%%%%%%%%%%%
\begin{center}
{\LARGE\bf Noncommutative Cohomological Field Theory}\\
{\LARGE\bf and GMS soliton}\\
\end{center}
%%%%%%%%%%%%%%%%%%%%%%%%%%%%%%%%%%%%%%%%%%%%%%%%%%%%%%%%%%%%%%%%%%%%%%%%%
\vspace{1.2cm}
Tomomi Ishikawa
\footnote{E-mail address: {\tt tomomi@theo.phys.sci.hiroshima-u.ac.jp}},
Shin-Ichiro~Kuroki
\footnote{E-mail address: {\tt kuroki@theo.phys.sci.hiroshima-u.ac.jp}} 
and Akifumi Sako
\footnote{E-mail address: {\tt sako@math.sci.hiroshima-u.ac.jp}}
\vspace{8mm}
\\
%%%%%%%%%%%%%%%%%%%%%%%%%%%%%%%%%%%%%%%%%%%%%%%%%%%%%%%%%%%%%%%%%%%%%%%%%
{\it Graduate School of Science, Hiroshima University,}\\
{\it 1-3-1 Kagamiyama, Higashi-Hiroshima 739-8526, Japan}
\vspace{1.8cm}
%%%%%%%%%%%%%%%%%%%%%%%%%%%%%%%%%%%%%%%%%%%%%%%%%%%%%%%%%%%%%%%%%%%%%%%%%
%%%%%%%%%%%%%%%%%%%%%%%%%%%%%%%%%%%%%%%%%%%%%%%%%%%%%%%%%%%%%%%%%%%%%%%%%
\begin{abstract}
We show that it is possible to construct a quantum
field theory that is invariant
under the translation of the noncommutative parameter $ \theta_{\mu \nu} $.
This is realized in a noncommutative cohomological field theory.
As an example, a noncommutative cohomological scalar field theory
 is constructed, and its partition function is calculated. The partition
 function is the Euler number of Gopakumar, Minwalla and Strominger (GMS)
 soliton space.
\end{abstract}
%%%%%%%%%%%%%%%%%%%%%%%%%%%%%%%%%%%%%%%%%%%%%%%%%%%%%%%%%%%%%%%%%%%%%%%%%
\end{center}
\end{titlepage}
%%%%%%%%%%%%%%%%%%%%%%%%%%%%%%%%%%%%%%%%%%%%%%%%%%%%%%%%%%%%%%%%%%%%%%%%%
\newpage
%%%%%%%%%%%%%%%%%%%%%%%%%%%%%%%%%%%%%%%%%%%%%%%%%%%%%%%%%%%%%%%%%%%%%%%%%
%%%%%%%%%%%%%%%%%%%%%%%%%%%%%%%%%%%%%%%%%%%%%%%%%%%%%%%%%%%%%%%%%%%%%%%%%
\section{Introduction}
%%%%%%%%%%%%%%%%%%%%%%%%%%%%%%%%%%%%%%%%%%%%%%%%%%%%%%%%%%%%%%%%%%%%%%%%%
\setcounter{equation}{0}
\setcounter{footnote}{0}
\renewcommand{\thefootnote}{\alph{footnote})}
%%%%%%%%%%%%%%%%%%%%%%%%%%%%%%%%%%%%%%%%%%%%%%%%%%%%%%%%%%%%%%%%%%%%%%%%%
Recently noncommutative geometry and noncommutative field theory
revived in the string theory~\cite{Nekrasov}. There are some correspondences
between
commutative and noncommutative geometry. For example in the gauge
theory, it was shown that the commutative theory with
some background field is equivalent to
the noncommutative  theory~\cite{witten}.
After that, a few kinds of nontrivial soliton in the noncommutative space
are
discovered. In the gauge theory, $U(1)$ instanton solution is
discovered by Nekrasov and Schwartz~\cite{N-S}, and in the scalar
theory a nontrivial solution is discovered by Gopakumar, Minwalla and
Strominger~\cite{GMS1}, which is called
GMS soliton. There is no
corresponding solution in the commutative space, i.e. the GMS
soliton is a specific solution in the noncommutative space.

In the noncommutative space, it is difficult to define the length or
metric. There are few examples, like a noncommutative torus case,
derived the differential geometry, e.g., connection and
curvature~\cite{connes}. But in general noncommutative space,
to define the Riemannian
geometry is difficult. What we can to do is to classify the geometry
to the extent of the algebric K-theory.
However, it is not enough to classify the
noncommutative space from a point of view of differential topology.
If there are some characteristic classes that do not varied under the
shift from commutative to noncommutative space, then they are useful for the
classification of spaces. For instance, in the noncommutative torus the
Euler number is independent of the noncommutative parameter $\theta$
~\cite{blackadar}. We expect that some other topological invariants
would be extended to noncommutative space and
independent of $\theta$.

The aim of this paper is to construct a quantum field
theory that is invariant under the transformation of the noncommutative
parameter $\theta^{\mu \nu} $. This parameter characterizes the
noncommutativity of spaces as
\begin{eqnarray}
\label{0.1}
[x^{\mu},x^{\nu}]=\frac{\theta^{\mu \nu}}{2\pi i},
\end{eqnarray}
where $x^\mu$ are the coordinates of the noncommutative space.
Noncommutative parameter independence of the theory means that the
partition function of the theory is independent of $\theta^{\mu \nu} $.
Cohomological field theory is nominated as such a theory.
We construct a cohomological field theory on the noncommutative space,
and we show that it is actually a usual cohomological field theory in the
$\theta^{\mu \nu} = 0$ limit.
This fact means that noncommutative space  succeed to some geometric or
topological information of commutative space.

Another purpose of this paper is to construct
a concrete example of the invariants under the transformation
of noncommutative parameter $\theta^{\mu \nu} $.
The discovery of the GMS solution is
one of the most important developments in recent work on
noncommutative field theory.
Therefore, we calculate the Euler number of the GMS soliton space
as the example of $\theta^{\mu \nu}$ independent
partition function.
We will see the relation between the GMS soliton and
commutative cohomological field theory.

 Our example of the noncommutative cohomological field theory is
a balanced scalar model that has two multiplets, the scalar and the
 vector. We consider the case that the potential
is general degree polynomial of the scalar field.
In the
 commutative space, the degree of the potential determines the structure
 of the vacuum. We will show that in the
 noncommutative space there is similar picture of vacuum structure. As a
result of
 investigation, the correspondence between the commutative limit
 $\theta\rightarrow 0$ and the noncommutative limit $\theta\rightarrow
 \infty$ is obtained.
Especially in the large $\theta$ limit, the potential term plays
 a dominant role, and there are the specific GMS
 solutions. A point of this paper is how we deal with the GMS solitons in
 quantum field theory.

This paper is organized as follows.
In section II, we will construct the $\theta^{\mu \nu}$ invariant
quantum field theory in general.
In section III, we construct a scalar model as a simple example.
Balanced topological theory will be used there.
It is necessary to introduce potential in topological scalar field
theory. Partition function is
calculated in both limits, commutative and noncommutative. In section IV,
we introduce the Morse theory on the noncommutative field theory, and
show that the Euler number of GMS soliton space is well-defined.
In the last section, we summarize and discuss  our result.
%%%%%%%%%%%%%%%%%%%%%%%%%%%%%%%%%%%%%%%%%%%%%%%%%%%%%%%%%%%%%%%%%%%%%%%%%
%%%%%%%%%%%%%%%%%%%%%%%%%%%%%%%%%%%%%%%%%%%%%%%%%%%%%%%%%%%
%%%%%%%%%%%%%%%%%%%%%%%%%%%%%%%%%%%%%%%%%%%%%%%%%%%%%%%%%%%%%%%%%%%%%%%%%
\section{General formalism}
\setcounter{equation}{0}
%%%%%%%%%%%%%%%%%%%%%%%%%%%%%%%%%%%%%%%%%%%%%%%%%%%%%%%%%%%%%%%%%%%%%%%%%
%%%%%%%%%%%%%%%%%%%%%%%%%%%%%%%%%%%%%%%%%%%%%%%%%%%%%%%%%%%
%%%%%%%%%%%%%%%%%%%%%%%%%%%%%%%%%%%%%%%%%%%%%%%%%%%%%%%%%%%%%%%%%%%%%%%%%
The aim of this section is to make a theory that is invariant
under the shift of the noncommutative parameter $\theta_{\mu \nu} $. We
discuss how we construct the cohomological field theory on the
noncommutative space.
%%%%%%%%%%%%2-1
\subsection{Cohomological field theory on noncommutative space}
The noncommutative parameter is defined in the commutation relation
Eq.(\ref{0.1}). We introduce the infinitesimal rescaling operator $\delta_s$
as follows.
%%%%%%%%%%%%2.3
\beq
\label{2.3}
   (1-\delta_s)
   [x^{\mu},x^{\nu}]
        \equiv [{x'}^{\mu},{x'}^{\nu}]
 =
    \frac{\theta^{\mu \nu}
    -\delta \theta^{\mu \nu}}{2\pi i}.
\eeq
This commutation relation is given by defining $\delta_s$ as
\beq
\label{2.4}
    {x'}^{\mu}
&=&
    x^{\mu} -\delta_s x^{\mu}, \\
        \delta_s x^{\mu}
&=&
    (\frac{1}{2}
     \delta \theta^{\mu \nu} (\theta^{-1})_{\nu \rho} )x^{\rho}.
\eeq
This transformation corresponds to ${x'}^{\mu}=\sqrt{\theta} x^{\mu}$  in
~\cite{GMS1}. We denote the inverse matrix of the transformation (\ref{2.4})
by
\beq
\label{2.7}
    J^{\mu}_\rho \equiv {\delta^{\mu}}_{\rho}+
\frac{1}{2} \delta \theta^{\mu \nu} (\theta^{-1})_{\nu \rho},
\eeq
Then, the integration measure and
the differential operator are transformed into
\beq
\label{2.8}
dx^D = \det {\mathbf{J}} {dx'}^{D}, \ \ \ \ \
\frac{\partial}{\partial x^{\mu}} =
(J^{-1} )_{\mu \nu} \frac{\partial}{{\partial x'}^{\nu},}
\eeq
where $ \det {\mathbf{J}}$ is the Jacobian. By (\ref{2.8}) the Moyal
product(see e.g.\cite{C-D-P}) is shifted as
\beq
\label{2.9}
(1-\delta_s) (\mbox{\Large $*$}_{\theta}) =
\delta_s (\exp(2 \pi i \overleftarrow{\partial}_{\mu}
(\theta-\delta \theta)^{\mu \nu} \overrightarrow{\partial}_{\nu}) )=
\mbox{\Large $*$}_{\theta -\delta \theta},
\eeq
because $\delta_s
(\overleftarrow{\partial }_{\mu} \theta^{\mu \nu}
\overrightarrow{\partial}_{\nu})= \overleftarrow{\partial }_{\mu} \delta
\theta^{\mu \nu} \overrightarrow{\partial}_{\nu}$.
Note that this transformation is just a rescaling of the coordinate, so
that any action and its partition function are not changed under this
transformation,
%%%%%%%%
\beq
\label{2.10}
S_{\theta}&=& \int dx^D {\cal L}(\mbox{\Large $*$}_{\theta}, \dmd )
\nonumber\\
          &=&    \int \det {\mathbf{J}} {dx'}^D
              {\cal L}(\mbox{\Large $*$}_{\theta-\delta \theta},
              (J^{-1} )^{\mu \nu} \frac{\partial}{\partial {x'}^{\nu}}),
\eeq
where ${\cal L}(\mbox{\Large $*$}_{\theta}, \dmd )$ is an explicit
description to emphasis that the products
of fields are the Moyal product and it contains
derivative terms in the lagrangian. For convenience, we will
often omit $\mbox{\Large $*$}_{\theta}$ when
we do not misunderstand. In the next step, we shift the noncommutative
parameter.
%%%%%%%%%%%%%%%%%
\beq
\label{2.11}
\theta \to \theta' =\theta + \delta \theta.
\eeq
This shift changes the action and the partition function in general, as
follows
%%%%%%%%%%%%%%%%%
\beq
\label{2.12}
S_{\theta'}=\int \det {\mathbf{J}} {dx'}^D
              {\cal L}(\mbox{\Large $*$}_{\theta},
              (J^{-1} )^{\mu \nu} \frac{\partial}{{\partial x'}^{\nu}}).
\eeq
Compared with (\ref{2.10}), the shift is regarded as a rescaling
without the Moyal product.

Contrary, our purpose is to construct a field theory invariant
under this shift.
Immediately we expect the cohomological field theory would be
an example since it is scale invariant
theory~\cite{B-B-R-T,2DY-M,TFT}. Cohomological field theory is
understood through several ways. Twisted SUSY is one of them, but
noncommutative SUSY is not adverted here~\cite{A-M-T,F-G-R,H-K-T}.
Meanwhile,
a geometrical point of view is closely studied in section IV.
The Lagrangian of cohomological field theory is BRST-exact.
We denote the BRST operator by $\hat{\delta}$, and generic bosonic fields
by $\phi_i$, which are sections of some vector bundle $s_a(\mbox{\Large
$*$}_{\theta}\phi_i)$. BRST operator is defined as
%%%%%%%%%%%%2.1
\beq
\label{2.1}
\hat{\delta} \phi_i = \psi_i, \ \ \ \  \hat{\delta} \psi_i =0, \nonumber\\
\hat{\delta} \chi^a= H^a, \ \ \ \  \hat{\delta} H^a =0,
\eeq
where $\phi_i$ and $H^a$ are bosonic, $\psi_i$ and $\chi^a$ are
fermionic fields.
Following the Mathai-Quillen formalism, the action of the cohomological
field theory is written as
%%%%%%%%%%%%2.2
\beq
\label{2.2}
V&=& \chi^a (i s_a + H_a ),\\
S_{\theta}&=& \int dx^D {\cal L}(\mbox{\Large $*$}_{\theta}, \dmd ) = \int
dx^D \hat{\delta} V
.
\eeq
The partition function is defined by
%%%%%%%%%%%%%2.2.1
\beq
\label{2.2.1}
Z_{\theta}= \int {\cal D}\phi \dps {\cal D}\chi {\cal D}H
\exp \left( -S_\theta \right).
\eeq
In the commutative space, the Mathai-Quillen formalism tells us that
the partition function gives a representation of the Euler number
of the space ${\cal M}=\{ s_a^{-1}(0) \} $.

This partition function is invariant under an infinitesimal
transformation which commute with the BRST transformation (\ref{2.1}).
%%%%%%%%%%%%%%%
\beq
\label{2.13}
\hat{\delta} \delta' &=& \pm \delta' \hat{\delta}, \nonumber\\
\delta' \ Z_{\theta} &=& \int {\cal D}\phi \dps {\cal D}\chi {\cal D}H
\ \ \delta' \left( -\int dx^D \hat{\delta} V \right) \ \exp \left( -S_\theta
\right)
\nonumber\\
&=&\int {\cal D}\phi \dps {\cal D}\chi {\cal D}H \ \
\hat{\delta} \left( -\int \delta' V \right) \ \exp \left( -S_\theta \right)
=0.
\eeq
The vacuum expectation value(VEV) of any BRST-exact observable is zero.
Note that the path integral measure is invariant under $\delta'$
transformation
 since every field has only one supersymmetric partner and
 the Jacobian is totally canceled.

 We introduce the $\theta$-shift operator as
 %%%%%%%%%%%%%%
 \beq
 \label{2.14}
 \delta_s x_{\mu} = -\delta_s x_{\mu} ,
 \ \ \ \delta_{\theta} \theta_{\mu \nu}=\theta_{\mu \nu} + \delta
\theta_{\mu \nu}.
 \eeq
Generally, it is possible to define $\delta_{\theta}$ to commute
with the BRST operator. Following (\ref{2.13}),
the partition function is invariant under this $\theta$-shift
and it means that the Euler number of the space ${\cal M}$ is
independent of the noncommutative parameter $\theta$.

In the end of this section, we list some general nature of this
partition function.
First, the Gaussian integral is formally defined and
an exact result can be given by 1-loop calculation.
Second, naively the commutative limit ($\theta \rightarrow 0$) is
given by removing potential terms without constant field, and
$\theta \rightarrow \infty $ limit is given by omitting
kinetic terms in the action, because the limit $\theta\rightarrow 0$
($\theta
\rightarrow \infty $) means ${\bf J}\rightarrow 0$ (${\bf J}\rightarrow
\infty$) in Eq.(\ref{2.12}).

In the next section, as an example of noncommutative cohomological
field theory, we investigate the balanced scalar model.
%%%%%%%%%%%%%%%%%%%%%%%%%%%%%%%%%%%%%%%%%%%%%%%%%%%%%%%%%%%%%%%%%%%%%%%%
%%%%%%%%%%%%%%%%%%%%%%%%%%%%%%%%%%%%%%%%%%%%%%%%%%%%%%%%%%%
\section{Balanced Scalar model}
\setcounter{equation}{0}

In this section, we study a concrete example of
noncommutative cohomological field theory as $\theta$-shift invariant
field theory. For simplicity, we use a real scalar model to construct
a cohomological field theory.
However, there is no nontrivial example under usual construction
of cohomological field theory. Therefore, we introduce the balanced
topological
field theory ~\cite{BTFT,lozano}.
This theory is used in the investigation of the Vafa-Witten theory
whose partition function is the sum of Euler number of the
zero section space ${\cal M}$ without a sign, in general
~\cite{DPS,sako,vafa-witten}.
When we calculate the Vafa-Witten theory, zero mode integration
is not essential because of no ghost number anomaly.
Contrary, in this paper, we carefully calculate the zero mode
integration to compare the commutative limit with the large $\theta$ limit.

\subsection{The action}
\begin{figure}[t]
 \begin{center}
   \psfragscanon
    \psfrag{P1}[][][3.0]{$\phi$}
    \psfrag{P2}[][][3.0]{$\psi$}
    \psfrag{C1}[][][3.0]{$\chi$}
    \psfrag{C2}[][][3.0]{$H$}
    \psfrag{C3}[][][3.0]{$0$}
    \psfrag{d+}[][][1.8]{$\hat\delta_+$}
    \psfrag{d-}[][][1.8]{$\hat\delta_-$}
    \psfrag{B1}[][][2.5]{$B^\mu$}
    \psfrag{B2}[][][2.5]{$\psi^\mu$}
    \psfrag{BC1}[][][2.5]{$\chi^\mu$}
    \psfrag{BC2}[][][2.5]{$H^\mu$}
    \scalebox{.4}{\includegraphics{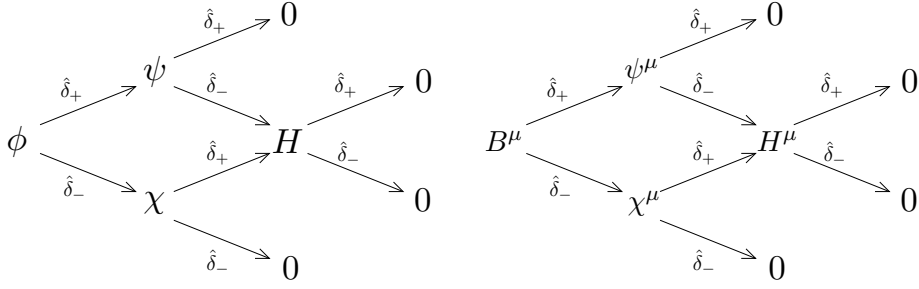}}
 \end{center}
\caption{BRST multiplets in the balanced scalar model.
$\delta_+(\delta_-)$ carries ghost number $+1(-1)$.}
\label{fig1}
\end{figure}

We construct the balanced scalar model.
The theory is composed of bosonic scalar fields $\phi$ and $H$,
fermionic scalar fields $\psi$ and $\chi$,
bosonic vector fields $B_{\mu}$ and $H_{\mu}$,
fermionic vector fields $\psi_{\mu}$ and $\chi_{\mu}$,
and every field is Hermitian field. We give the BRST
transformations (Fig.\ref{fig1}) as
\begin{eqnarray}
 &&\hat{\delta}_+\phi = \psi ,\;\hat{\delta}_-\phi = \chi ,\nonumber\\
 &&-\hat{\delta}_-\psi = \hat{\delta}_+\chi = H ,\nonumber\\
 &&\hat{\delta}_+\psi = \hat{\delta}_-\chi
  = \hat{\delta}_+H = \hat{\delta}_-H =0 ,\nonumber\\
 &&\hat{\delta}_+B^{\mu} = \psi^{\mu} ,
  \;\hat{\delta}_-B^{\mu} = \chi^{\mu} ,\nonumber\\
 &&-\hat{\delta}_-\psi^{\mu} = \hat{\delta}_+\chi^{\mu} = H^{\mu}
,\nonumber\\
 &&\hat{\delta}_+\psi^{\mu} = \hat{\delta}_-\chi^{\mu}
  = \hat{\delta}_+H^{\mu} = \hat{\delta}_-H^{\mu} = 0 ,
\end{eqnarray}
where $\hat{\delta}_+$ and $\hat{\delta}_-$ satisfy the following relations,

\begin{equation}
 \hat{\delta}_+^2 = \hat{\delta}_-^2 =0 ,\;\;\;
  \hat{\delta}_+\hat{\delta}_-+\hat{\delta}_-\hat{\delta}_+ = 0 .
\end{equation}
The action of the balanced scalar model is
\begin{equation}
 S=\int dx^D \sqrt{g}\mathcal{L},
\end{equation}
where the Lagrangian $\mathcal{L}$ is given by an exact form with a
ghost-number zero functional $\mathcal{F}$ as
\begin{equation}
 \mathcal{L}=i\hat{\delta}_+\hat{\delta}_-\mathcal{F} .
\end{equation}
In our case, the functional $\mathcal{F}$ is chosen as
\begin{equation}
 \mathcal{F} = B^{\mu}\partial_{\mu}\phi-i\chi^{\mu}\psi_{\mu}-i\chi\psi
  +V(\phi,B^{\rho}) ,
\end{equation}
where $V(\phi,B^{\rho})$ is a potential.
Therefore the Lagrangian $\mathcal{L}$ becomes
\begin{eqnarray}
 \mathcal{L}&=&i\hat{\delta}_+\hat{\delta}_-\mathcal{F}\nonumber\\
            &=&i\hat{\delta}_+\left\{\chi^{\mu}\partial_{\mu}\phi
       +B^{\mu}\partial_{\mu}\chi-i\chi^{\mu}H_{\mu}-i\chi H
       +\left(\chi\frac{\delta}{\delta\phi}\right)V(\phi,B^{\rho})
       +\left(\chi^{\mu}\frac{\delta}{\delta B^{\mu}}\right)
        V(\phi,B^{\rho})\right\}\nonumber\\
            &=&i\hat{\delta}_+\left[\chi\left\{-\partial_{\mu}B^{\mu}
       +\frac{\delta}{\delta\phi}V(\phi,B^{\rho})-iH\right\}
       +\chi^{\mu}\left\{\partial_{\mu}\phi
               +\frac{\delta}{\delta
B^{\mu}}V(\phi,B^{\rho})-iH_{\mu}\right\}
       \right]\nonumber\\
    &=&\mathcal{L}_B+\mathcal{L}_F ,
\end{eqnarray}
where $\mathcal{L}_B \ (\mathcal{L}_F)$ is the bosonic (fermionic) part
of the Lagrangian,
\begin{eqnarray}
 \mathcal{L}_B &=& iH\left\{-\partial_{\mu}B^{\mu}
+\frac{\delta}{\delta\phi}V(\phi,B^{\rho})-iH\right\}
+iH^{\mu}\left\{\partial_{\mu}\phi
        +\frac{\delta}{\delta B^{\mu}}V(\phi,B^{\rho})-iH_{\mu}\right\} ,\\
 \mathcal{L}_F &=& i\chi\left\{\partial_{\mu}\psi^{\mu}
        -\left(\psi\frac{\delta}{\delta\phi}\right)\frac{\delta}{\delta\phi}
         V(\phi,B^{\rho})
        -\left(\psi^{\mu}\frac{\delta}{\delta B^{\mu}}\right)
\frac{\delta}{\delta\phi}V(\phi,B^{\rho})\right\}\nonumber\\
               & & -i\chi^{\mu}\left\{\partial_{\mu}\psi
        +\left(\psi\frac{\delta}{\delta\phi}\right)
\frac{\delta}{\delta B^{\mu}}V(\phi,B^{\rho})
        +\left(\psi^{\nu}\frac{\delta}{\delta B^{\nu}}\right)
         \frac{\delta}{\delta B^{\mu}}V(\phi,B^{\rho})\right\} .
\end{eqnarray}
%%%%%%%%%%%%%%%%%%%%%%%%%%%%%%%%%%%%%%%%%%%%%%%%%%%%%%%%%%%%%%%%%%%%%%%%
%%%%%%%%%%%%fig:BRST multiplet%%%%%%%%%%%%%%%%%%%%%%%%%%%%%%%
%%%%%%%%%%%%%%%%%%%%%%%%%%%%%%%%%%%%%%%%%%%%%%%%%%%%%%%%%%%%%
%%%%%%%%%%%%%%%%%%%%%%%%%%%%%%%%%%%%%%%%%%%%%%%%%%%%%%%%%%%%%%%%%%%%%%%%
%%%%%%%%%%%%%%%%%%%%%%%%%%%%%%%%%%%%%%%%%%%%%%%%%%%%%%%%%%%%%%%%%%%%%%%%
%%%%%%%%%%%%%%%%%%%%%%%%%%%%%%%%%%%%%%%%%%%%%%%%%%%%%%%%%%%%%%%%%%%%%%%%%
Here we consider the only case that the potential is separated form as
\begin{equation}
 V(\phi,B^{\rho})=V(\phi)+\frac{1}{2}B_{\mu}B^{\mu} .
\end{equation}
As a result of taking this potential, $B^{\mu}$ has only zero solution
in the large $\theta$ limit, and this is necessary condition to
investigate the GMS soliton space. If there is non-trivial  $B^{\mu}$
solution, then the moduli space is changed from GMS soliton moduli space
and we are not interested in such a case.
$\mathcal{L}_B$ and $\mathcal{L}_F$, therefore, become
\begin{eqnarray}
 \mathcal{L}_B &=& iH\left\{-\partial_{\mu}B^{\mu}
        +\frac{\delta V(\phi)}{\delta\phi}-iH\right\}
+iH^{\mu}(\partial_{\mu}\phi+B_{\mu}-iH_{\mu}),\label{nc4-1}\\
 \mathcal{L}_F &=& i\chi\left\{\partial_{\mu}\psi^{\mu}
        -\psi\frac{\delta^2V(\phi)}{\delta\phi^2}\right\}
-i\chi^{\mu}(\partial_{\mu}\psi+\psi_{\mu}).\label{nc4-2}
\end{eqnarray}

Note that every product of fields should be defined by replacing the
normal product by the Moyal product when we consider the
noncommutative field theory.
%%%%%%%%%%%%%%%%%%%%%%%%%%%%%%%%%%%%%%%%%%%%%%%%%%%%%%%%%%%%%%%%%%%%%%%%%
%%%%%%%%%%%%%%%%%%%%%%%%%%%%%%%%%%%%%%%%%%%%%%%%%%%%%%%%%%%%%%%%%%%%%%%%%
\subsection{Commutative limit $\theta \rightarrow 0$}
We consider the balanced scalar model in 2-dimensional flat
noncommutative space.
One can get the theory on noncommutative space by changing the ordinary
product into the star (Moyal) product $\mbox{\Large $\ast$}_{\theta}$.
Although noncommutativity is represented by the noncommutative parameter
$\theta$ in the star product, this parameter is absorbed
by the rescaling,
\begin{eqnarray}
      x^{\mu}
    &\rightarrow
     &\sqrt{\theta}x^{\mu} ,\nonumber\\
       \mbox{\Large $\ast$}_{\theta}
           &\rightarrow
            &\mbox{\Large $\ast$}_{\theta=1} .
\end{eqnarray}
However, the action is changed as follows:
\begin{eqnarray}
 S_B\hspace{-3mm}&=&\hspace{-3mm}\int d^2x\theta\left[iH\ast
         \left\{-\frac{\partial_{\mu}B^{\mu}}{\sqrt{\theta}}
                 +\frac{\delta V(\ast\phi)}{\delta\phi}-iH\right\}
+iH^{\mu}\ast\left(\frac{\partial_{\mu}\phi}{\sqrt{\theta}}
         +B_{\mu}-iH_{\mu}\right)\right],\\
 S_F\hspace{-3mm}&=&\hspace{-3mm}\int d^2x\theta\left[i\chi\ast\left\{
         \frac{\partial_{\mu}\psi^{\mu}}{\sqrt{\theta}}
                 -\psi\frac{\delta^2V(\ast\phi)}{\delta\phi^2}\right\}
-i\chi^{\mu}\ast\left(
\frac{\partial_{\mu}\psi}{\sqrt{\theta}}
+\psi_{\mu}\right)\right],
\end{eqnarray}
where $S_B\ (S_F)$ is the bosonic (fermionic) part of the action and
$\mbox{\Large $\ast$}$
means $\mbox{\Large $\ast$}_{\theta=1}$ implicitly.\par
Let us consider the $\theta\rightarrow 0$ limit and calculate its
partition function, which is compared with the
$\theta\rightarrow\infty$ limit in the next subsection.
The partition function is calculated as
\begin{equation}
 \mathcal{Z}=\int\mathcal{D}\phi\mathcal{D}\chi\mathcal{D}\psi
  \mathcal{D}\chi^{\mu}\mathcal{D}\psi^{\mu}\mathcal{D}B^{\mu}\mathcal{D}
  H\mathcal{D}H^{\mu}\exp(-S_B-S_F).
\end{equation}
The part of the action which contributes to the non-zero modes of
$\phi$ and $H$ is
\begin{equation}
 \int d^2x
  \sqrt{\theta}i(H^{\mu}\ast\partial_{\mu}\phi-H\ast\partial_{\mu}B^{\mu}).
\end{equation}
We perform the integration of the non-zero modes of $H^{\mu}$ and $B^{\mu}$,
and it yields delta functionals,
\begin{equation}
 \delta(\sqrt{2\pi\theta}\partial_{\mu}\phi)
  \delta(\sqrt{2\pi\theta}\partial_{\mu}H).
\end{equation}
Therefore, for $\phi$ and $H$, only zero modes integral remains.
Next, the part of the action which contributes to the non-zero modes of
$\chi$ and $\psi$ is
\begin{equation}
 \int d^2x
  \sqrt{\theta}i(\chi\ast\partial_{\mu}\psi^{\mu}
  -\chi^{\mu}\ast\partial_{\mu}\psi) .
\end{equation}
Integration of non-zero modes of $\chi,\psi$ and $\chi^{\mu},\psi^{\mu}$
yields the factor
\begin{equation}
 [\det(i\sqrt{\theta}\partial_{\mu})]
  _{\chi_{\scriptscriptstyle\emptyset}\psi^{\mu}}
  [\det(i\sqrt{\theta}\partial_{\mu})]
  _{\chi^{\mu}\psi_{\scriptscriptstyle\emptyset}}
  ,
\end{equation}
where the fields with the subscript ${}_{\emptyset}$ denote non-zero modes.
Therefore, for $\chi$ and $\psi$, only zero modes remain and
the partition function $\mathcal{Z}$ becomes
\begin{eqnarray}
 \mathcal{Z}&=&
  \frac{[\det(i\sqrt{\theta}\partial_{\mu})]_
  {\chi^{\mu}\psi_{\scriptscriptstyle\emptyset}}}
  {[\det(\sqrt{2\pi\theta}\partial_{\mu})]_
  {H^{\mu}\phi_{\scriptscriptstyle\emptyset}}}
  \frac{[\det(i\sqrt{\theta}\partial_{\mu})]_
  {\chi_{\scriptscriptstyle\emptyset}\psi^{\mu}}}
  {[\det(\sqrt{2\pi\theta}\partial_{\mu})]_
  {H_{\scriptscriptstyle\emptyset}B^{\mu}}}\nonumber\\
 &&\hspace{1cm}\times
  \int dyd\chi_{\scriptscriptstyle 0}d\psi_{\scriptscriptstyle 0}
  \mathcal{D}\chi^{\mu}\mathcal{D}\psi^{\mu}
  \mathcal{D}B^{\mu}dH_{\scriptscriptstyle 0}
  \mathcal{D}H^{\mu}\exp(-S^{\prime}_B-S^{\prime}_F) ,
  \label{EQ:001}
\end{eqnarray}
where the fields with the subscript ${}_0$ denote zero modes and
the variable $y$ denotes the zero mode of $\phi$ (this is just a real
constant number) and
\begin{eqnarray}
 S^{\prime}_B&=&\int d^2x \theta\left[H_{\scriptscriptstyle 0}\left\{
    H_{\scriptscriptstyle 0}+i\frac{\delta V(y)}{\delta y}\right\}
    +H^{\mu}\ast(H_{\mu}+iB_{\mu})\right],\\
 S^{\prime}_F&=&\int d^2x\theta i
  \left[-\chi_{\scriptscriptstyle 0}\psi_{\scriptscriptstyle 0}
   \frac{\delta^2V(y)}{\delta y^2}
   -\chi^{\mu}\ast\psi_{\mu}\right].
\end{eqnarray}
The factors in front of the integral in Eq.(\ref{EQ:001}) cancel each
other and only $[\det(2\pi)]^{-1}
_{H^{\mu}\phi_{\scriptscriptstyle\emptyset}}$ remains.
$H^{\mu},B^{\mu}$ and the zero mode of $H$ can be integrated out,
and it yields the factor,
\begin{equation}
  \frac{1}{[v\theta/\pi]_{H_{\scriptscriptstyle 0}}^{\frac{1}{2}}}
  \frac{1}{[\det(\theta/\pi)]_{H^{\mu}\phi_{\scriptscriptstyle 0}}
  ^{\frac{1}{2}}}
  \frac{1}{[\det(\theta/4\pi)]_{H_{\scriptscriptstyle 0}B^{\mu}}
  ^{\frac{1}{2}}},
\end{equation}
where $v$ is the volume of space-time. This volume is infinity but this
will be canceled out by $\phi$, $\chi$ and $\psi$
zero modes integration, later.
Integration of $\chi^{\mu},\psi^{\mu}$ yields the factor
$[\det(i\theta)]_{\chi^{\mu}\psi^{\mu}}$ ,
and the partition function $\mathcal{Z}$ becomes
\begin{equation}
 \mathcal{Z}=\sqrt{\frac{\pi}{v\theta}}\det(-1)
  \int dyd\chi_{\scriptscriptstyle 0}d\psi_{\scriptscriptstyle 0}
  \exp(-S^{\prime\prime}_B-S^{\prime\prime}_F),
\end{equation}
where
\begin{eqnarray}
 S^{\prime\prime}_B&=&\int d^2x
  \frac{\theta}{4}\left\{\frac{\delta V(y)}{\delta y}\right\}^2 ,\\
 S^{\prime\prime}_F&=&-\int d^2x
  \theta i\chi_{\scriptscriptstyle 0}\psi_{\scriptscriptstyle 0}
  \frac{\delta^2V(y)}{\delta y^2} .
\end{eqnarray}
Before integrating $y$, we expand $\delta V(y)/\delta y$ as
\begin{equation}
 \frac{\delta V(y)}{\delta y}
  =y\left.\frac{\delta^2 V(y)}{\delta y^2}\right|_{y=y_c}+\mathcal{O}(y^2) ,
\end{equation}
where $y_c$ is a point of the extrema of the potential,
\begin{equation}
 \left.\frac{\delta V(y)}{\delta y}\right|_{y=y_c}=0 ,
\end{equation}
and we should sum up all $y_c$ in the calculation of $\mathcal{Z}$:
\begin{equation}
 \mathcal{Z}=\sqrt{\frac{\pi}{v\theta}}\det(-1)
  \sum_{y_c}\int dyd\chi_{\scriptscriptstyle 0}d\psi_{\scriptscriptstyle 0}
  \exp(-S^{\prime}) ,
\end{equation}
where
\begin{equation}
 S^{\prime}=v\theta\left[
  \frac{1}{4}\left\{y\left.\frac{\delta^2V(y)}{\delta y^2}\right|_{y=y_c}
  \right\}^2-i\chi\psi\left.\frac{\delta^2V(y)}{\delta y^2}\right|_{y=y_c}
  \right] .
\end{equation}
Note that these zero modes do not depend on $x^\mu$, then the volume $v$ is
factorized out. By integrating the zero modes of $\phi,\chi$ and $\psi$,
we get the partition function,
\begin{eqnarray}
 \mathcal{Z}&=&\sqrt{\frac{\pi}{v\theta}}\det(-1)\sum_{y_c}
  \frac{iv\theta\left.\frac{\delta^2V(y)}{\delta y^2}\right|_{y=y_c}}
  {\left\{\frac{v\theta}{4\pi}\left.
  \frac{\delta^2V(y)}{\delta y^2}\right|_{y=y_c}^2\right\}^{\frac{1}{2}}}
\nonumber\\
 &=&2\pi i\det(-1)
  \sum_{y_c}\mbox{sgn}\left[\left.\frac{\delta^2V(y)}{\delta y^2}
  \right|_{y=y_c}\right] .\label{EQ:002}
\end{eqnarray}
The factor in front of $\sum_{y_c}$ is removable as a normalizing factor.

Generally, the potential $V(\phi)$ is a polynomial of the scalar field
$\phi$:
\begin{equation}
 V(\phi)=b_0+b_1\phi+\frac{b_2}{2!}\phi^2+\cdots+\frac{b_m}{m!}\phi^m
  \;\;\;\;(m\geqslant 2) ,
\end{equation}
and so the result Eq.(\ref{EQ:002}) becomes
\begin{eqnarray}
 \mathcal{Z}=\begin{cases}
      \mbox{sgn}[b_m] & :m \ {\rm \ is\ even\ number},\\
      \;\;\;\;0       & :m \ {\rm \ is\ odd\ number}.
     \end{cases}\label{EQ:003}
\end{eqnarray}
In the limit $\theta\rightarrow 0$, it seems that the kinetic term plays a
dominant role, but the effect of the potential term survives.
By this effect,
degree of the polynomial completely determine the partition function
as cyclic form of Eq.(\ref{EQ:003}).
%%%%%%%%%%%%%%%%%%%%%%%%%%%%%%%%%%%%%%%%%%%%%%%%%%%%%%%%%%%%%%%%%%%%%%%%%
%%%%%%%%%%%%%%%%%%%%%%%%%%%%%%%%%%%%%%%%%%%%%%%%%%%%%%%%%%%%%%%%%%%%%%%%%
\subsection{Noncommutative limit $\theta\rightarrow\infty$}
\setcounter{equation}{0}
%%%%%%%%%%%%%%%%%%%%%%%%%%%%%%%%%%%%%%%%%%%%%%%%%%%%%%%%%%%%%%%%%%%%%%%%%
In the strong noncommutative limit $\theta \rightarrow \infty$, the
terms that have derivatives are effectively ignored, and the remaining
terms which are potential and mass terms in the balanced scalar model
determine the field configuration. In the noncommutative space, there
is a specific field configuration, that is
GMS soliton~\cite{GMS1}. This soliton is the solution of the field
equation, and there are infinite solutions.
Hence, the partition function is sum of contributions from
infinite vacuum states.

In the large $\theta$ limit, the balanced scalar model
Eqs.(\ref{nc4-1},\ref{nc4-2}) is
written as
\begin{eqnarray}
 S=S_B+S_F,
\end{eqnarray}
where
\begin{eqnarray}
  S_B
   &=&\int d^2x\theta
       \left[H\ast
               \left\{\frac{\delta V(\ast\phi)}{\delta\phi}-H\right\}
+H^{\mu}\ast
                      \left(B_{\mu}-H_{\mu}\right)\right],\label{nc5-1}\\
   S_F
    &=&\int d^2x\theta
        \left[-i\chi\ast\left(
         \psi\frac{\delta}{\delta\phi}\right)
          \frac{\delta V(\ast\phi)}{\delta\phi}
          +i\chi^{\mu}\ast\psi_{\mu}\right].\label{nc5-2}
\end{eqnarray}
After integrating over the fields $H$, $H^\mu$, the action is
\begin{eqnarray}
  S=\int d^2x\theta
     \left[
      \frac{1}{4}\frac{\delta V(*\phi)}{\delta \phi}
       \frac{\delta V(*\phi)}{\delta \phi}
       -i\chi*\left(\psi\frac{\delta}{\delta\phi}\right)
   \frac{\delta V(*\phi)}{\delta \phi}
   +\frac{1}{4}B^\mu *B_\mu
    +i\chi^\mu *\psi_\mu
      \right].
\end{eqnarray}
Integration of the fields $B^\mu$, $\chi^\mu$ and $\psi_\mu$ is simply
performed, and the partition function is written as
\begin{equation}
   \mathcal{Z}
   =\int\mathcal{D}\phi
         \mathcal{D}\chi
          \mathcal{D}\psi e^{-S},
\end{equation}
where the action is
\begin{eqnarray}
  S=\int d^2x\theta
     \left[
      \frac{1}{4}\frac{\delta V(*\phi)}{\delta \phi}
       \frac{\delta V(*\phi)}{\delta \phi}
       -i\chi*\left(\psi\frac{\delta }{\delta \phi}\right)
         \frac{\delta V(*\phi)}{\delta \phi}
          \right].
\end{eqnarray}
In the large $\theta$ limit, the derivative terms are irrelevant, and the
potential terms dominates{\footnote {Actually kinetic term integral
is survived for next order integral but it yield nothing because of the BRST
symmetry.}}. The field configuration is determined by
the form of the potential. In particular the stationary field configuration
is
obtained by solving the field equation:
\begin{eqnarray}
    \frac{\delta V(*\phi)}{\delta \phi}=0.
\end{eqnarray}
 In the calculation of the partition function,
we should treat these GMS soliton as the stationary points, and the quantum
fluctuation is a perturbation from the GMS soliton. In the following
section,
we discuss the treatment of the GMS soliton in the partition function,
and understand the GMS soliton from a topological view point.\\

%%%%%%%%%%%%%%%%%%%%%%%%%%%%%%%%%%%%%%%%%%%%%%%%%%%%%%%%%%%%%%%%%%%%%%%%%
%
%%%%%%%%%%%%%%%%%%%%%%%%%%%%%%%%%%%%%%%%%%%%%%%%%%%%%%%%%%%%%%%%%%%%%%%%%
We consider the $\phi ^{m+2}$ potential,
\begin{eqnarray}
     V(\phi)
       =\{\phi{\rm -polynomial\  of }\ (m+2)\ {\rm degree}\},
\end{eqnarray}
%%%%%%%%%%%%%%%%%%%%%%%%%%%%%%%%%%%%%%%%%%%%%%%%%%%%%%%%%%%%%%%%%%
%%%%%%%%%%%%%%%%%%%%%%%%%%%%%%%%%%%%%%%%%%%%%%%%%%%%%%%%%%%%%%%%%%%%%%%%%
\begin{figure}[t]
  \begin{center}
   \psfragscanon
    \psfrag{X1}[][][2.0]{$\frac{\delta V}{\delta\phi}$}
    \psfrag{X2}[][][1.8]{$v_0=0$}
    \psfrag{X3}[][][1.8]{$v_1$}
    \psfrag{X4}[][][1.8]{$v_2$}
    \psfrag{X5}[][][1.8]{$v_3$}
    \psfrag{X6}[][][1.8]{$v_{m-1}$}
    \psfrag{X7}[][][1.8]{$v_m$}
    \psfrag{Y1}[][][2.0]{$\phi$}
     \scalebox{.5}{\includegraphics{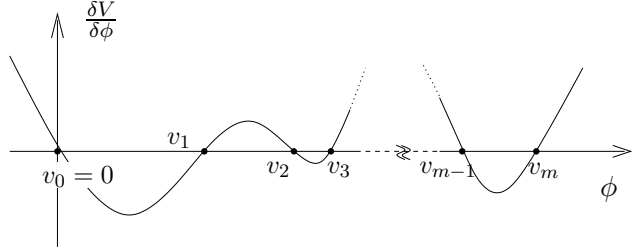}}
  \end{center}
\caption{$\frac{\delta V}{\delta\phi}$, $b_m>0$ and $m$ is odd.}
\label{fig2}
\end{figure}
%%%%%%%%%%%%%%%%%%%%%%%%%%%%%%%%%%%%%%%%%%%%%%%%%%%%%%%%%%%%%%%%%%
%%%%%%%%%%%%%%%%%%%%%%%%%%%%%%%%%%%%%%%%%%%%%%%%%%%%%%%%%%%%%%%%%%%%%%%%%
and the field equation takes the following factorized form,
\begin{eqnarray}
     \frac{\delta V}{\delta\phi}
   &=&
    b_m \phi*
         \left(\phi -v_1\right)*
          \left(\phi -v_2\right)*
           \cdots*
            \left(\phi -v_m\right). \label{l-1}
\end{eqnarray}
Where we assume the $v_i$ are real constant numbers with
$v_1<v_2<\cdots < v_{m-1} <v_{m}$.
The GMS soliton is given by
\begin{eqnarray}
    \phi_{GMS}=\lambda_i {\bf P}_i,
\end{eqnarray}
where ${\bf P}_i$ is a projection, and satisfies the idempotent
relation ${\bf P}_i*{\bf P}_j=\delta_{ij}{\bf P}_i$. The coefficient
$\lambda_i$ is determined by Eq.(\ref{l-1}). In the Moyal plane,
this projection is given by the Laguerre polynomial
${\scriptstyle 2\left(-1\right)^ie^{-x^2}{\rm
L}_i\left(2x^2\right)}$~\cite{GMS1}.
It is possible to chose concrete representation of
corresponding Weyl mapped projection operator,
for example
\begin{eqnarray}
     \hat{\bf P}_i=|i\rangle\langle i|,
\end{eqnarray}
where $|i\rangle$ is the number representation base~\cite{D-N,kajiura}.
The Weyl mapped GMS soliton is written as
\begin{eqnarray}
    \hat\phi_{GMS}=\lambda_i \hat{\bf P}_i.
\end{eqnarray}
We substitute the GMS solution to $\phi$ of Eq.(\ref{l-1})
\begin{equation}
   \left.\frac{\delta V(\hat\phi)}{\delta\hat\phi}\right|_{\scsc GMS}
   =
    b_m \lambda_i
         \left(\lambda_i-v_1\right)
          \left(\lambda_i -v_2\right)
           \cdots
            \left(\lambda_i -v_m\right){\bf P}_i=0. \label{l-2}
\end{equation}
The solutions of Eq.(\ref{l-2}) are given by
\begin{eqnarray}
      \lambda_i=0, ~v_1, ~v_2, ~\cdots , ~v_{m}.
\end{eqnarray}
These solutions are coefficients of the GMS soliton.
The general GMS solution is a linear combination of the projections,
\begin{eqnarray}
\label{GMS}
    \hat\phi _{\scriptscriptstyle GMS}
  &=&
      \sum_i \lambda_i \hat{\bf P}_i \nonumber\\
  &=&
     v_1\sum_{i\in {\bf S}_1}\hat{\bf P}_i
      +v_2 \sum_{i\in {\bf S}_2}\hat{\bf P}_i
       +\cdots
        + v_m \sum_{i\in {\bf S}_m}\hat{\bf P}_i\\
&=&
     v_1\hat{\bf P}_{\scsc {\bf S}_1}
      +v_2 \hat{\bf P}_{\scsc {\bf S}_2}
       +\cdots
        + v_m \hat{\bf P}_{\scsc {\bf S}_m},\nonumber
\end{eqnarray}
where ${\bf S}_A~\left(~A\in \{ 1,2,\cdots, m\} \right)$ is a set of the
indices of the projections, and it is defined
as  if the coefficient of a projection $\hat{\bf P}_i$ is $v_A$
in (\ref{GMS}) then  the index ``$i$'' belongs
to ${\bf S}_A$ .
For example, if a GMS soliton takes the form as
$\hat\phi_{\scsc GMS}=v_1\left(\hat{\bf P}_1+\hat{\bf P}_2\right)
+v_2\left(\hat{\bf P}_0\right)$, then ${\bf S}_1=\{1,~2\}$ and
${\bf S}_2=\{0\}$. We define $\hat{\bf P}_{\scsc {\bf S}_A}
\equiv\sum_{i\in \hat{\bf S}_A}{\bf P}_i$.
For $A\neq B ~\left(~A, B\in \{ 1,2,\cdots, m\} \right)$ the sets
${\bf S}_A$ and ${\bf S}_B $ are disjoint each other
${\bf S}_A\cap {\bf S}_B=\o$, and the projections are orthogonal,
 $\hat{\bf P}_{\scsc {\bf S}_A}
\cdot \hat{\bf P}_{\scsc {\bf S}_B}=0$. In the commutative
space, the field equation Eq.(\ref{l-1}) has only constant solutions.
On the contrary, in the noncommutative space, there exist GMS solitons that
are operators for Hilbert space of quantized coordinate, and this fact
make new solutions with
 linear combination of projections. The indices of
the projection are not bounded above, then the number of the element of
${\bf S}_A$ (${\rm rank}{\bf S}_A$) is allow to be infinite. In the
following, we introduce
a cut-off $N$ of total number of the ${\rm rank}{\bf S}_A$, and then
there are $N$ projections $\left\{\hat{\bf P}_0,~\hat{\bf P}_1,~\hat{\bf
P}_2,~\cdots, ~\hat{\bf P}_{N-1}\right\}$.
After the calculation we take the limit $N\rightarrow\infty$. \\

Next step, we perform Gaussian integral around each vacuum.
Note that, in the commutative space the number of the solution is
$\left(m+1\right)$, on the contrary in the noncommutative space there
are $\left(m+1\right)^N$ solutions. In the commutative space,
we can take the quantum fluctuation from the vacua of the
finite number of simply constant solutions of Eq.(\ref{l-1}).
On the contrary, there are infinite GMS
solitons in noncommutative space.
Let us take the quantum fluctuation around each of the GMS
solitons as $\phi = \phi|_{\scsc GMS}+\phi_q$. We expand the
Lagrangian around the specific GMS soliton,
\begin{eqnarray}
   {\cal L}_{\bf S}
  &=&
    \hspace*{-1mm}
      \frac{1}{4}
       \left(\phi_q \frac{\delta }{\delta \phi}
        \right)
         \left.\frac{\delta V(*\phi)}{\delta \phi}
          \right|_{\scriptscriptstyle GMS}
          \hspace{-5mm}
          *\left(\phi_q \frac{\delta }{\delta \phi}
            \right)
             \left.\frac{\delta V(*\phi)}{\delta \phi}
              \right|_{\scriptscriptstyle GMS}
          \hspace{-5mm}
       -i\chi*\left(\psi\frac{\delta }{\delta \phi}
          \right)
           \left.\frac{\delta V(*\phi)}{\delta \phi}
            \right|_{\scriptscriptstyle GMS}\hspace{-7mm}.\hspace{5mm}
\end{eqnarray}
The subscript ${\bf S}$ of ${\cal L}$ means that the Lagrangian is
expanded around the GMS soliton specified by ${\bf S}$ that is the
family of $\left\{{\bf S}_1, {\bf S}_2,\cdots,{\bf S}_m
\right\}$ {\footnote {Rank $k$ single GMS soliton space is infinite
dimensional moduli space of solution and it is concretely described as
the coset space $\frac{U(N)}{U(k)U(N - k)}$ \cite{G-H-S}. In our case, GMS
solution space contains not only single soliton but
also multi-solitons that have any rank solutions. The moduli space
specified by ${\bf S}$ is$\frac{U(N)}{U(n_0)U(n_1)\cdots U(N-1)}$, where
$n_i$ is the rank of ${\bf S}_{i}$.}}. On the Moyal plane, the field
operators $\hat\phi_q$, $\hat\chi$ and
$\hat\psi$ are expanded as
\begin{eqnarray}
    \hat\phi_q
  &=&
     \sum_{i,j=0}^{\infty} \phi_{ij}|i\rangle\langle j|,\label{l-3}\\
      \hat\chi
    &=&\sum_{i,j=0}^{\infty} \chi_{ij}|i\rangle\langle j|,\nonumber \\
        \hat\psi
      &=&\sum_{i,j=0}^{\infty} \psi_{ij}|i\rangle\langle j| \nonumber .
\end{eqnarray}
The field operators $\hat\phi_q$, $\hat\chi$ and $\hat\psi$
are Hermitian operators,
\begin{eqnarray}
      \phi_q^\dag =\phi_q,
      ~\chi^\dag =\chi,
       ~\psi^\dag=\psi.
\end{eqnarray}
We rewrite the field such as $\phi_{ij}=\phi_{ij}^R+i\phi_{ij}^I$
$(\phi_{ij}^R,~\phi_{ij}^I\in {\bf R})$.
Then the hermitian conditions need that real (imaginary) parts are
symmetric (anti-symmetric)
\begin{eqnarray}
    \left\{\begin{array}{ll}
         \phi_{ij}^R=\phi_{ji}^R, &\phi_{ij}^I=-\phi_{ji}^I,\\
         \chi_{ij}^R=\chi_{ji}^R, &\chi_{ij}^I=-\chi_{ji}^I,\\
         \psi_{ij}^R=\psi_{ji}^R, &\psi_{ij}^I=-\psi_{ji}^I.
     \end{array}\right.
\end{eqnarray}
In the operator picture, the integration in the action is replaced by
the trace over the field operators. The lagrangian is always
described as the Weyl mapped lagrangian
\begin{eqnarray}
     S_{\bf S}
  =
     {\rm Tr}\hat{\cal L}_{\bf S}
  =
      S_B^{\bf S}+S_F^{\bf S},
\end{eqnarray}
where $S_B^{\bf S}$ and $S_F^{\bf S}$ are given by
\begin{eqnarray}
     S_B^{\bf S}
 &=&
        \frac{1}{4}\theta ~{\rm Tr}~
           \left(\hat{\phi}_q\frac{\delta}{\delta \hat{\phi}}
           \right)
            \left.\frac{\delta V(\hat{\phi})}{\delta\hat{\phi}}
             \right|_{\scsc GMS}\hspace{-2mm}
           \left(\hat{\phi}_q\frac{\delta}{\delta \hat{\phi}}
            \right)
             \left.\frac{\delta V(\hat{\phi})}{\delta\hat{\phi}}
              \right|_{\scsc GMS},
               \\
     S_F^{\bf S}
&=&
        -i\theta~{\rm Tr}~
            \hat{\chi}
             \left(\hat{\psi}\frac{\delta}{\delta\hat{\phi}}
              \right)
         \left.\frac{\delta V(\hat{\phi})}{\delta\hat{\phi}}
          \right|_{GMS}.
\end{eqnarray}
Firstly, we consider the bosonic part of the action. The second
derivative of the potential is
\begin{eqnarray}
    \lefteqn{
       \left(\hat\phi_q \frac{\delta }
        {\delta \hat\phi}
        \right)
         \left.\frac{\delta V(\hat\phi)}{\delta \hat\phi}
          \right|_{\scsc GMS}}\hspace{10mm} \nonumber\\
&=&
       \left(\hat\phi_q \frac{\delta }{\delta \hat\phi}
        \right)
         \left\{\left. b_m \hat\phi
         \left(\hat\phi -v_1\right)
          \left(\hat\phi -v_2\right)
           \cdots
            \left(\hat\phi -v_m\right)\right\}\right|_{\scsc GMS}
\nonumber\\
&=&
b_m\left[
       \hat\phi_q
        \left(\hat\phi-v_1\right)
          \left(\hat\phi-v_2\right)
          \cdots
          \left(\hat\phi-v_m\right)\right.\nonumber\\
  &&
        \hspace*{3mm}+\hat\phi
        \hat\phi_q
         \left(\hat\phi-v_2\right)
         \cdots
         \left(\hat\phi-v_m\right)\nonumber\\
   &&
        \hspace*{3mm}+\hat\phi
        \left(\hat\phi-v_1\right)
         \phi_q
          \cdots
           \left(\hat\phi-v_m\right)\nonumber\\
    &&
        \hspace*{3.0mm} \vdots\nonumber\\
   &&
        \hspace*{3mm}+\left.\hat\phi
        \left(\hat\phi-v_1\right)
         \left(\hat\phi-v_2\right)
         \cdots
         \hat\phi_q \right]_{\hat\phi=\hat\phi_{GMS}}.\label{2nd-deri}
\end{eqnarray}
We put the GMS soliton into Eq.(\ref{2nd-deri}), and we get
\begin{eqnarray}
    \lefteqn{
       \left(\hat\phi_q \frac{\delta }{\delta \hat\phi}
        \right)
         \left.\frac{\delta V(\hat\phi)}{\delta \hat\phi}
          \right|_{\scsc GMS}}\hspace{10mm} \nonumber\\
  &=&
      b_m
        \sum_{A=0}^{m}
         \sum_{\scsc B=A}^{\scsc m}
          \prod_{\scsc i=0}^{\scsc A-1}
           \left(v_B-v_i\right)
               \sum_{\scsc C=0}^{\scsc A}
           \prod_{\scsc j=A+1}^{\scsc m}
            \left(v_C-v_j\right)
       \hat{\bf P}_{\scsc {\bf S}_B}
        \hat\phi
            \hat{\bf P}_{\scsc {\bf S}_C}
              \nonumber\\
 &=&
     b_m\sum_{\scsc A=0}^{\scsc m}
         \prod_{\scsc i=0}^{\scsc A-1}\left(v_A-v_i\right)
          \prod_{\scsc j=A+1}^{\scsc m}\left(v_A-v_j\right)
           \hat{\bf P}_{\scsc {\bf S}_A}
            \hat\phi_q
             \hat{\bf P}_{\scsc {\bf S}_A}
              \nonumber\\
 &&
      +b_m \hspace{-5mm}\sum_{\stackrel{\mbox {$\scsc A>B$}}
             {\mbox{$\scsc A,B\in\{0,\cdots,r\} $}}}
           \sum_{k=B}^A
           \prod_{\scsc i=0}^{\scsc k-1}
            \left(v_A-v_i\right)
            \prod_{\scsc j=k+1}^{\scsc m}
             \left(v_B-v_j\right)
          \hat{\bf P}_{\scsc {\bf S}_A}
           \hat\phi_q
            \hat{\bf P}_{\scsc {\bf S}_B},\label{2nd-deri-GMS}
\end{eqnarray}
where we define the zero$^{th}$ solution and the zero$^{th}$ projection
by
\begin{eqnarray}
   v_0\equiv 0,
  \hspace{2mm}
   \hat{\bf P}_{\scsc {\bf S}_0}
    \equiv
     \hat{\bf 1}
     -\hat{\bf P}_{\scsc {\bf S}_1}
      -\hat{\bf P}_{\scsc {\bf S}_2}
       -\cdots
       - \hat{\bf P}_{\scsc {\bf S}_r},
\end{eqnarray}
and we introduce the following symbols for convenience
\begin{eqnarray}
    \prod_{\scsc i=0}^{\scsc -1}\left(v_0-v_i\right)
     \equiv 1,
      \hspace{2mm}
       \prod_{\scsc j=m+1}^{\scsc m}\left(v_m-v_j\right)
        \equiv 1.
\end{eqnarray}
The coefficient of the cross term
$\hat{\bf P}_{\scsc {\bf S}_A} \hat\phi_q {\bf P}_{\scsc {\bf S}_B}
\left(A> B\right)$ in Eq.(\ref{2nd-deri-GMS}) is
\begin{eqnarray}
 \sum_{k=B}^A
  \prod_{i=0}^{k-1}
   \left(v_A -v_i\right)
    \prod_{j=k+1}^{m}
   \left(v_B -v_j\right).\label{coeff1}
\end{eqnarray}
Eq.(\ref{coeff1}) is written as
\begin{eqnarray}
 \lefteqn{\sum_{k=B}^A
  \prod_{i=0}^{k-1}
   \left(v_A -v_i\right)
    \prod_{j=k+1}^{m}
   \left(v_B -v_j\right)}\hspace{10mm}\nonumber\\
       &=&
       \left(v_A-v_0\right)
       \left(v_A-v_1\right)
        \cdots
       \left(v_A-v_{B-1}\right)\nonumber\\
    &&
       \hspace*{5mm}\times
        \left(v_{A+1}-v_B\right)
         \left(v_{A+2}-v_B\right)
          \cdots
            \left(v_{m-1}-v_B\right)
        \left(v_{m}-v_B\right) { \$}_{A-B}, \nonumber\\
\end{eqnarray}
where $\$_{A-B}$ is defined as
\begin{eqnarray}
   \$_{A-B}
      &\equiv&
       \left(v_B -v_{B+1}\right)
     \left(v_B -v_{B+2}\right)
      \cdots
      \left(v_B -v_{A}\right)\nonumber\\
    &&
    +\left(v_A -v_{B}\right)
     \left(v_B -v_{B+2}\right)
      \cdots
      \left(v_B -v_{A}\right)\nonumber\\
  &&\hspace*{1mm}  \vdots \nonumber\\
    &&+\left(v_A -v_{B}\right)
     \left(v_{B+1} -v_{A}\right)
      \cdots
      \left(v_{A-1} -v_{A}\right).
\end{eqnarray}
We can see $\$_{A-B}$ always vanishes for any set of
$v_i$(see Appendix D). Then the
cross term does not appear. Finally the remaining
terms are closed in each set ${\bf S}_A$ such as
\begin{eqnarray}
       \left(\hat\phi_q \frac{\delta }{\delta \hat\phi}
        \right)
         \left.\frac{\delta V(\hat\phi)}{\delta \hat\phi}
          \right|_{\scsc GMS}
&=&
     b_m\sum_{\scsc A=0}^{\scsc m}
       \prod_{\scsc i=0}^{\scsc A-1}\left(v_A-v_i\right)
        \prod_{\scsc j=A+1}^{\scsc m}\left(v_A-v_j\right)
         \hat{\bf P}_{\scsc {\bf S}_A}
          \hat\phi_q
           \hat{\bf P}_{\scsc {\bf S}_A}.
\end{eqnarray}
Then the bosonic part of the action is written as
\begin{eqnarray}
     S_B^{\bf S}
 &=&   \frac{1}{4}\theta
        \sum_{\scsc A=0}^{\scsc m}
         \left(b_m\prod_{\scsc i=0}^{\scsc A-1}\left(v_A-v_i\right)
          \prod_{\scsc j=A+1}^{\scsc m}\left(v_A-v_j\right)\right)^2
          {\rm Tr}\left[\hat{\bf P}_{\scsc {\bf S}_A}
            \hat\phi_q
             \hat{\bf P}_{\scsc {\bf S}_A}
              \hat\phi_q \right].   \label{l-6}
      \end{eqnarray}
For the fermionic part, the calculation is the same as the bosonic part.
$\left(\hat\psi \frac{\delta }{\delta \hat\phi}
        \right)
         \frac{\delta V(\hat\phi)}{\delta \hat\phi}
          $
is given as
\begin{eqnarray}
\left(\hat\psi \frac{\delta }{\delta \hat\phi}
        \right)
         \left.\frac{\delta V(\hat\phi)}{\delta \hat\phi}
          \right|_{\scsc GMS}
&=&
     b_m\sum_{\scsc A=0}^{\scsc m}
       \prod_{\scsc i=0}^{\scsc A-1}\left(v_A-v_i\right)
        \prod_{\scsc j=A+1}^{\scsc m}\left(v_A-v_j\right)
        \hat{\bf P}_{\scsc {\bf S}_A}
          \hat\psi
          \hat{\bf P}_{\scsc {\bf S}_A}. \nonumber\\
\end{eqnarray}
Then the fermionic part of the action is written as
\begin{eqnarray}
      S_F^{\bf S}
&=&
    -i\theta b_m\sum_{\scsc A=0}^{\scsc m}
       \prod_{\scsc i=0}^{\scsc A-1}\left(v_A-v_i\right)
        \prod_{\scsc j=A+1}^{\scsc m}\left(v_A-v_j\right)
        {\rm Tr}
          \left[
           \hat\chi
            \hat{\bf P}_{\scsc {\bf S}_A}
             \hat\psi
              \hat{\bf P}_{\scsc {\bf S}_A}
               \right]. \label{l-8}
\end{eqnarray}
We substitute ${\bf P}_{\scsc {\bf S}_A}=\sum_{i\in{\bf S}_A }{\bf P}_i$
into
Eqs.(\ref{l-6}, \ref{l-8}), and use Eq.(\ref{l-3}), then we get the action,
\begin{eqnarray}
     S_{\bf S}
   &=&
     S_B^{\bf S}+S_F^{\bf S} \nonumber\\
   &=&
      \sum_{\scsc A=0}^{\scsc m}
      ~\left\{
        \frac{1}{4}\theta
         \sum_{i,j\in{\bf S}_{A}}
          \left(b_m \zeta_A\right)^2
           \phi_{ij}
            \phi_{ji}
     -i\theta
        \sum_{i,j\in{\bf S}_{A}}
      b_m\zeta_A
           \chi_{ij}
            \psi_{ji}\right\},
\end{eqnarray}
where we define
\begin{eqnarray}
      \zeta_A
       \equiv
        \prod_{\scsc k=0}^{\scsc A-1}\left(v_A-v_k\right)
         \prod_{\scsc l=A+1}^{\scsc m}\left(v_A-v_l\right),
          \hspace{2mm}
          A\in\{0,1,2,\cdots, m\}.
\end{eqnarray}
Note that the $\phi_{ij}$, $\chi_{ij}$ and $\psi_{ij}$ are the c-numbers.
Using the result in Appendix C, we can estimate both the partition function
in the operator representation and in the commutative field representation
with the Moyal product, but the results have no difference.
Here, we perform the integral in the operator representation.
Then the path-integral becomes simply an integral of
real number (real Grassmann number)
.
\begin{eqnarray}
    {\cal Z}_{\bf S}
 &=&
       \int
       {\cal D}\chi
        {\cal D}\psi
         {\cal D}\phi
          ~e^{-S_{\bf S}}\nonumber\\
 &=&
       \int\prod_{ A=0}^{m}
            \prod_{\stackrel{\mbox {$\scriptscriptstyle m>n$}}
                   {\mbox{$\scriptstyle i,j\in {\bf S}_A $}}}
             \left( \frac{d\phi_{ij}}{\sqrt{4\pi\theta}}\right)
             d\chi_{ij}d\psi_{ij}
                      ~\exp\{-S_B^{\bf S}-S_F^{\bf S}\}\label{measure} \\
 &=&
   \prod_{ A=0}^{m}
     \left[
          \sqrt{b_m \zeta_A}
   \right]^{-n_A^2}
    \left[
           ~b_m \zeta_A~
       \right]^{n_A^2} ,\nonumber
\end{eqnarray}
and we get
\begin{eqnarray}
   {\cal Z}_{\bf S}
        &=&
      \prod_{A=0}^{m}
       \left[
      ~{\rm sgn}
         \left(~b_m \zeta_A~\right)
         ~\right]^{n_A}.
\end{eqnarray}
Here $n_A$ is the number of the elements belong to the set ${\bf S}_A$, and
we
call this number the ``rank'' of ${\bf S}_A$~;~$n_A={\rm rank}{\bf
S}_A$. The  noncommutative parameter
$\theta$ contained under integral measure $d\phi$
in the second line of (\ref{measure}) comes from the integral of $H$.
The path integral of the
fields $\phi_{ij}$, $\chi_{ij}$ and
$\psi_{ij}$ with indices ${\scriptstyle (i\in {\bf S}_A, j\in
{\bf S}_B , A\neq B)}$ is done with the weight of
 the kinetic terms, which contributes
``1'' in the partition function. The number of the
GMS solitons specified by ${\bf S}=\{ {\bf S_1 ,S_2, \cdots, S_m}\}$ is
${\scriptstyle \frac{N!}{n_0!n_1!\cdots n_m!}}$,
where $n_i= {\rm rank}{\bf S_i}$.
Therefore the total partition function
that includes all the GMS solitons is given by
\begin{eqnarray}
  {\cal Z}_{Total}
     &=&\lim_{N\rightarrow\infty}\sum_{\bf S} {\cal Z}_{\bf S}\nonumber\\
     &=&\lim_{N\rightarrow\infty}
     \sum_{\stackrel{\mbox {$\scsc n_0, \cdots, n_m=0$}}
                {\mbox{$\scsc n_0+\cdots+n_m=N$}}}^N
      {\frac{N!}{n_0!n_1!\cdots n_m!}}\prod_{A=0}^{m}
       \left[
      ~{\rm sgn}
         \left(~b_m \zeta_A~\right)
         ~\right]^{n_A} \label{l-10}\\
    &=&\lim_{N\rightarrow\infty}
       \left(
          {\rm sgn}\left[b_m\right]\right)^N
           \left(
            {\rm sgn}\zeta_0
           +{\rm sgn}\zeta_1
           +\cdots
            +{\rm sgn}\zeta_m\right)^N.
\end{eqnarray}
Under the same condition of $v_0 < v_1 < \cdots <v_m$ as
previous section (see fig.2 too),
the sign of $\zeta_A$ is simple form,
\begin{eqnarray}
{\rm sgn}\left(~\zeta_A~\right)=(-1)^{m-A},
           \hspace{2mm}
          A\in\{0,1,2,\cdots, m\}.
\end{eqnarray}
Then the partition function is
\begin{eqnarray}
    {\cal Z}_{Total}
   &=&
       \lim_{N\rightarrow\infty}
        \left(
        {\rm sgn}
          \left[b_m \right] \right)^N
           \left(-1 \right) ^m
   (~\underbrace{
            1
           -1
           +1
           -1
           +\cdots
            }_{m+1}~)^N \nonumber\\
   &=&
       \left\{\begin{array}{ll}
         \lim_{\scsc N\rightarrow\infty}
        \left(
        {\rm sgn}
          \left[b_m \right] \right)^N& : m\ {\rm  is\ even\ number},\\
0 &  :m\ {\rm is\ odd\ number }.
    \end{array}
     \right.
\end{eqnarray}
\makebox \\

Note that through a simple calculation, the partition function
Eq.(\ref{l-10}) is
rewritten as,
\begin{eqnarray}
   {\cal Z}_{Total}
 &=&
      \lim_{N\rightarrow\infty}
       \left[~{\rm sgn}(b_m)~\right]^N(-1)^m \nonumber\\
 &&\hspace*{6mm}      \times  \sum_{n_p=0}^{N}(-1)^{n_p}
        \frac{N!}{(N-n_p)!n_p!}
       \left[(m+1)/2~\right]^{n_p}
                      \left(\left[m/2~\right]
                     +1\right)^{N-n_p} ,
\end{eqnarray}
where the bracket$[\cdots]$ is the Gauss's notation. The integer $n_p$ is
the
number of the negative solutions, and it is the sum of $n_{2i+1}$.
This number $n_p$ is related to
the Morse theory. This is discussed in the next section.
%%%%%%%%%%%%%%%%%%%%%%%%%%%%%%
%%%%%%%%%% section 4
%%%%%%%%%%%%%%%%%%%%%%%%%%%%%%
\section{Euler number of GMS soliton space}
\setcounter{equation}{0}
%%%%%%%%%%%%%%%%%%%%%%%%%%%%%%%%%%%%%%%%%%%%%%%%%%%%%%%%%%%%%%%%%%
%%%%%%%%%%%%%%%%%%%%%%%%%%%%%%
In this section we discuss the properties of
the partition function in the point of view of
quantum geometry.\\

%%%%%%%%%%%%2-1
\subsection{Noncommutative Mathai-Quillen formalism}
In the beginning, we study the partition function
with Mathai-Quillen formalism.\\

As in the usual cohomological field theory,
$\psi$ is a tangent vector to the zero-section, i.e.
the tangent vector to the solution space,
$ \{ \phi, B_{\mu} | s^a (\phi , B_{\mu}, \theta )=0 \}={\cal M}$.
Note that the vanishing theorem asserts that
the zero-section space is identified
with the GMS solution space. In our theory, since $B^{\mu}B_{\mu}=0$
has only a trivial solution, the zero section space is identified with
$\{ \phi | \frac{\delta V}{\delta \phi} = 0 \}$.
A condition like the vanishing theorem appears in general in the
balanced topological field theory.\\

In the commutative limit ($\theta \rightarrow 0$),
the relevant terms of the bosonic action are
%%%%%%%%%%%%
\beq
\label{4.1}
| \dmd \phi |^2 + \left| \frac{\delta V}{\delta \phi} \right|^2.
\eeq
So the lowest energy solutions (true vacua) are constant that satisfy
$\frac{\delta V}{\delta \phi}=0$.
The number of the solutions is $m+1$, that is the exponent
of $\frac{\delta V}{\delta \phi}$ in the commutative limit.
Hence, there is no problem in defining the Euler number
of the space ${\cal M}$ which consists of $m+1$ isolate points
as $\pm\sum_{k=0}^m(-1)^k$.
Contrary, in the large $\theta$ limit, the number of GMS soliton
solutions is infinite , that is rank of $K_0$.
Generally, this is  difficult case
to define the Euler number of ${\cal M}$.
However, as we saw in the previous sections,
the partition function is invariant under the shift of $\theta$.
Finally,
the Euler number of the moduli space is invariant under the noncommutative
deformation of $\frac{ \delta V}{\delta \phi}=0$.\\

Gaussian integral in the noncommutative field theories is not
familiar in general.
But in this case, it is well-defined by the supersymmetry.
As we see at first, the result of the path integral in the
strong noncommutative limit does not contradict to the commutative limit.
This is an evidence of validity of the Gaussian integral
of the noncommutative field theory.\\
Note that the large $\theta$ limit is not the strong coupling limit.
We have to consider two parameters, the noncommutative parameter $\theta$
and the coupling constant $g$ multiplying overall.
In order to get exact results by perturbative calculations,
it is necessary for $g$ to approach to zero.\\
In the cohomological field theory, the partition function does not
depend on an overall parameter like this $g$.

We summarize the previous sections by taking account
of the fact that the partition
function of cohomological field theory can be regard as the Euler number of
${\cal M}$ by Mathai-Quillen formalism~\cite{blau}.
\newtheorem{theorem}{Theorem}[section]
%%%%%%%%%%%%%
\begin{theorem}[$\theta$-shift invariants]
The partition functions $Z_{\theta}$ of the noncommutative
cohomological field theories are invariant under the shift
of the noncommutative parameter $\theta$.
\beq
\label{4.1.1}
\frac{\delta}{\delta \theta} Z_{\theta} = 0.
\ \ \ \ \ \ \ \ \ \ \ \ \ \ \ \ \ \ \ \ \ \ \ \rule{3mm}{3mm}
\eeq
\end{theorem}

%%%%%%%%%%%%%
\begin{theorem}[Euler number of GMS solution space]
On the Moyal plane, when the GMS solution space for real scalar
field $\phi$ is \\
\beq
\label{4.2}
{\cal M}_{m}= \{ \phi |\
 b_m ( \cdots ( \underbrace{\phi \mbox{\Large $*$}_\theta (\phi -v_1)
)\mbox{\Large $*$}_\theta (\phi -v_2))\cdots \mbox{\Large $*$}_\theta
(\phi -v_m)}_{m+1} =0,
\ b_m>0 ~\},
\eeq
and any two of $v_i$ are different $(v_i \neq v_j for i \neq j)$,
then the Euler number
of the space ${\cal M}_{m}$ is given by
\beq
\label{4.3}
\chi_{m} =
\left\{
\begin{array}{ccc}
1 &: {\rm m\ is\ even\  number} &\makebox{} \\
0 &: {\rm m\ is\ odd\ number.} &
\ \ \ \ \ \ \ \ \ \ \ \ \ \ \rule{3mm}{3mm}
\end{array} \right.
\eeq
\end{theorem}
%%%%%%%%%%%%%%%%%%%%%%%%%%%%%
\subsection{Noncommutative Morse Theory}
In this subsection, we see that the noncommutative cohomological field
theory
can be regarded as the noncommutative Morse theory~\cite{Milnor,witten}.\\

In the commutative limit,
the partition function of our theory is given
by the Hessian of the $\frac{\delta^2 V}{\delta \phi^2}  $
as we saw in the section III.
Especially, the zero mode of $\dmd \phi=0$ is just a real
constant number, then the determinant of the Hessian
 at the critical point $p$ is determined by the sign of the second
derivative
of the real function at $p$, where the critical point $p$ is the
solution of $\left. \frac{d\ V(x)}{dx} \right|_p=0 \  \ (x \in {\bf R}) $.
We denote the number of negative eigenvalues of the Hessian
as $n_p$. In the commutative limit $n_p$ is either 0 or 1, so the
partition function is written by
%%%%%%%%%%%%%%%
\beq
\label{4.5}
 Z=
  \sum_p{\frac{\det\frac{\delta^2 V(p)}{\delta p^2}}{\left|
  \det \frac{\delta^2 V(p)}{\delta p^2} \right|} } = \sum_p (-1)^{n_p}
 =
 \left\{
\begin{array}{cc}
~1~ :& m {\rm \ is \ even \  number},\\
~0~ :& m {\rm \ is \ odd \  number}.
\end{array} \right.
\eeq
By the fundamental theorem of Morse theory, this is the Euler number
of the isolated points $\{p \}$ and this result is consistent with
the result of
applying Mathai-Quillen formalism to
cohomological field theory.

On the other hand, the partition function is
determined by the determinant of
the operator valued Hessian $\frac{\delta^2 V(\phi )}{\delta \phi^2}$
in the large $\theta$ limit.
The critical points are operators too, i.e.GMS solitons are
critical points and they exist infinitely at each $p$.
But for each critical point, we can estimate
$\det \!\!\left(\frac{\delta^2 V(\phi
)}{\delta \phi^2}\right) $.
Now we introduce the Morse index $M_{n_p}$.
$M_{n_p}$ is defined as the number of the GMS
solitons satisfying the condition
that Hessian consisting of the GMS solitons has $n_p$ negative eigenvalues.
As we saw in the section III and the Appendix, the $n_{p}$
 is the number of projection operators ${\bf P}_i$
whose coefficient belong to
$p_{-}$ in the GMS
 solution. Here we define $p_{-}$ and $p_+$ as the set of
 the critical points that
 satisfy
\beq
\label{4.5.1}
\begin{array}{c}
\left. \frac{\delta^2 V(p)}{\delta p^2}\right|_{p_{-}}< 0, \\
\left.  \frac{\delta^2 V(p)}{\delta p^2}\right|_{p_{+}}> 0.
\end{array}
\eeq
In the case of positive $b_m$, the critical points are,
$p_-=\left\{v_0,~v_2,~\cdots,~v_m\right\}$, and
$p_+=\left\{v_1,~v_3,~\cdots,~v_{m-1}\right\}$(see Fig.\ref{fig2}).
The number of the $p_{-}$ elements is $[(m+1)/2]$, and
the number of the $p_{+}$
is $[m/2]+1$. Then the number of
combinations that $n_p$ projections is combined with  $[(m+1)/2]$
points of $p_{-}$ in the soliton is
  $[(m+1)/2]^{n_p}$. When the total number of projections is fixed by
$N$,  then the remaining $(N-n_p)$ projections are combined with $p_{+}$ and
its number of combinations is $([m/2]+1)^{(N-n_p)}$.
Then the Morse index is given by
%%%%%%%%
\beq
\label{4.6}
M_{n_p}(m,N) =
[(m+1)/2]^{n_p}([m/2]+1)^{(N-n_p)}
\cdot \frac{N!}{(N-n_P)! n_p !} .
\eeq

$M_{n_p}$ is divergent in the limit of $N \rightarrow \infty$.
But we can define the Euler number of the isolate
GMS soliton solutions as the fundamental theorem of  Morse theory,
%%%%%%%
\beq
\label{4.7}
\chi_{m}  &=& \sum_{GMS} (-1)^{n_p} = \sum_{n_p}(-1)^{n_p}M_{n_p}
= (-[(m+1)/2] + [m/2]+1)^N \\
&=& \left\{
\begin{array}{cc}
~1~ :& m {\rm \ is \ even \  number},\\
~0~ :& m {\rm \ is \ odd \  number}.
\end{array} \right.
\eeq
This is a consistent result with the section III and the Mathai-Quillen
formalism. Therefore we can conclude that the noncommutative cohomological
field theory makes it possible to generalize the Morse theory
for the noncommutative field theory.
%%%%%%%%%%%%%%%%%%%%%%%%
%%%%%%%%%%%%%%%%%%%%%%%%%%%%
%%%%% section 5
%%%%%%%%%%%%%%%%%%%%%%%%%%%%
\section{Conclusion and Discussion}
\setcounter{equation}{0}

We have studied the noncommutative cohomological field theory.
Especially, the balanced scalar model is investigated carefully.
A couple of theorems is provided.
First, the partition function is invariant under the shift of
the noncommutative parameter. Second, the Euler number of the
GMS soliton space on the Moyal plane is calculated and it is ``1'' for
the scalar potential with even degree and ``0'' for odd
degree\footnote
{Note that the potential in original GMS paper correspond
to $\frac{\delta V(\phi)}{\delta \phi}$ in our theory. So the degree
of potential is shifted from $m$ to $m+1$.}.
In general, Mathai-Quillen formalism tells us that
 the partition function of cohomological field theory
is the Euler number of solution space ${\cal M}$ when the space is
commutative.
We expect that the noncommutative case is the same as the commutative case.
Indeed, as we saw in the previous section, it
is possible to identify  the partition function of the
noncommutative cohomological field theory with the Euler number
as a result of the fundamental theorem of Morse theory, that is
extended to the noncommutative field theory. In this paper, we saw this
relation in a scalar model, but it is expected that there is
no obstacle to applying general cases.

It is possible to use our method for more complex models.
For example, we can estimate the Euler number of the moduli space of
 instantons on noncommutative $\textbf{R}^4$.
In that case, the partition function is the Euler number of
the instanton moduli space, and there are some new moduli space
of the new instanton like Nekrasov-Schwarz~\cite{N-S,IKS}. For other
example, we can change the base manifold to noncommutative torus.
In that case, we will have to use the Powers-Rieffel projection
for calculation in the strong noncommutative
limit~\cite{kajiura,rieffel}. Another point of view, we should study
other types of noncommutativity. For example noncommutative parameter is
locally defined type, a fuzzy sphere type and so on.
To study such various cases is important to
construct the local geometry of noncommutative spaces. They are left as
future works.

Most of the geometric nature of the noncommutative space
is still unknown.
But, as we saw, it is likely that
some kinds of nature
of commutative space are succeeded by noncommutative space.
We will have to study huge amount of them to throw light on the
noncommutative
geometry.
\\

\hspace*{-6mm}
{\Large\bf  Acknowledgment}\vspace*{-2mm}\\

We are grateful to H.Kanno for helpful suggestions and
observations and a critical reading of the manuscript.
We also would like to thank H.Moriyoshi
for valuable discussion.
A.Sako is supported by JSPS Research Fellowships for
Young Scientists.
\vspace{-10mm}
%%%%%%%%%%%%%%%%%%%%%%%%%%%%%%%%%%%%%%%%%%%%%%%%%%%%%%%%%%%%%%%%%%%%%%%%%
\appendix
\vspace*{20mm}
\hspace*{-6mm}{\Huge\bf  Appendix.}
\renewcommand{\theequation}{\Alph{section}. ~\arabic{equation}}
%%%%%%%%%%%%%%%%%%%%%%%%%%%%%%%%%%%%%%%%%%%%%%%%%%%%%%%%%%%%%%%%%%%%%%%%%
\section{Large $\theta$ limit:$\phi^3$ potential}
\setcounter{equation}{0}
%%%%%%%%%%%%%%%%%%%%%%%%%%%%%%%%%%%%%%%%%%%%%%%%%%%%%%%%%%%%%%%%%%%%%%%%%
In this appendix, we show calculations  in the large $\theta$ limit
for simple examples of the balanced
scalar model. One example is the $\phi^3$ potential, and
another is the $\phi^4$ potential that is discussed in Appendix B.

We consider the cubic potential,
\begin{eqnarray}
     V(\phi)
       =b_0 +b_1\hat\phi
                +\frac{1}{2}b_2 \phi*\phi
                 +\frac{1}{3!}b_3 \phi*\phi*\phi.
\end{eqnarray}
The field equation is written as
\begin{eqnarray}
     \frac{\delta V}{\delta\phi}
   &=&b_1+b_2 \phi +\frac{1}{2}b_3 \phi*\phi \nonumber\\
   &=&\frac{1}{2}b_3(\phi-\alpha)*(\phi-\beta),
\end{eqnarray}
where we put
\begin{eqnarray}
 \alpha, ~\beta =-(b_2/b_3)\pm \sqrt{(b_2/b_3)^2-2(b_1/b_3)}.
\end{eqnarray}
We consider the case both $\alpha$ and $\beta$ are real numbers.
By the redefinition of the field $\phi$ by a translation,
\begin{eqnarray}
     \phi '\equiv \phi-\alpha,
\end{eqnarray}
the field equation is written as
\begin{eqnarray}
   \frac{\delta V}{\delta\phi}
   =\frac{1}{2}b_3 \phi '*\left(\phi ' -(\beta-\alpha)\right)=0.
\end{eqnarray}
The GMS solution is given by
\begin{eqnarray}
    \hat\phi '_{GMS}=\lambda_i \hat{\bf P}_i,
\end{eqnarray}
where the projection operator is $\hat{P}_i=|i\rangle\langle i|$.
We put the GMS solution to the field equation,
\begin{eqnarray}
     \frac{\delta V(\hat\phi)}{\delta\hat\phi}
     =\frac{1}{2}b_3 \lambda_i(\lambda_i-(\beta-\alpha))\hat{\bf P}_i =0.
\end{eqnarray}
The solution is
\begin{eqnarray}
      \lambda_i=0, ~\lambda_i=\beta-\alpha\equiv v.
\end{eqnarray}
The general GMS solution is a linear combination of projections with
coefficients $\lambda_i$,
\begin{eqnarray}
    \hat\phi_{\scriptscriptstyle GMS}
   =v\sum_{i\in {\bf S}}\hat{\bf P}_i
  =v\hat{\bf P_S}.
\end{eqnarray}
The bosonic part of the action is
\begin{eqnarray}
 S_B^{\bf S}
 &=&
    -\frac{1}{4}\theta~{\rm Tr}~
                   \left(\hat{\phi}_q\frac{\delta}{\delta \hat{\phi}}
              \right)
               \left.\frac{\delta V(\hat{\phi})}{\delta\hat{\phi}}
                \right|_{GMS}
           \left(\hat{\phi}_q\frac{\delta}{\delta \hat{\phi}}
            \right)
             \left.\frac{\delta V(\hat{\phi})}{\delta\hat{\phi}}
              \right|_{GMS} .
\end{eqnarray}
The linear terms of expansion of $V(\hat\phi)$ around the GMS soliton are
\begin{eqnarray}
    \left(\hat\phi_q\frac{\delta}{\delta \hat\phi}
     \right)
      \left.\frac{\delta V(\hat\phi)}{\delta\hat\phi}
       \right|_{GMS}
   &=&
    \left(\hat\phi_q \frac{\delta}{\delta\hat\phi}
     \right)
      \left[\frac{b_3}{2}\hat\phi (\hat\phi-v)
       \right]_{\phi=\phi_{GMS}}   \nonumber\\
   &=&
     \frac{b_3}{2}
      \left\{\hat\phi_q(\hat\phi_{\scsc GMS}-v)
      +\hat\phi_{\scsc GMS}\hat\phi_q
        \right\}                     \nonumber\\
    &=&
      \frac{b_3v}{2}
       \left\{-\hat\phi_q (\hat{\bf 1}-\hat{\bf P_S})+\hat{\bf
P_S}\hat\phi_q \right\}.
\end{eqnarray}
Hence, we get
\begin{eqnarray}
 S_B^{\bf S}
    &=&
      -\frac{1}{4}\theta
        \left(\frac{b_3v}{2}\right)^2
           {\rm Tr}
           \left[
             \hat{\phi}_q \left(\hat{\bf 1}-{\bf \hat{P}_S}\right)
              \hat{\phi}_q \left(\hat{\bf 1}-{\bf \hat{P}_S}\right)
             +{\bf \hat{P}_S}\hat{\phi}_q {\bf \hat{P}_S}\hat{\phi}_q
           \right].
\end{eqnarray}
The fermionic part of the action is
\begin{eqnarray}
     S_F^{\bf S}
  &=&
        -i\theta~{\rm Tr}
            ~\hat{\chi}
             \left(\hat{\psi}\frac{\delta}{\delta\hat{\phi}}
              \right)
         \left.\frac{\delta V(\hat{\phi})}{\delta\hat{\phi}}
          \right|_{GMS}.
\end{eqnarray}
As is the same as the bosonic part, the leading terms of quantum field are
given by
\begin{eqnarray}
    \left(\hat\psi\frac{\delta}{\delta \hat\phi}
     \right)
      \left.\frac{\delta V(\hat\phi)}{\delta\hat\phi}
       \right|_{GMS}
       &=&
      \frac{b_3v}{2}
       \left\{
       -\hat\psi (\hat{\bf 1}-\hat{\bf P_S})
        +\hat{\bf P_S}\hat\psi \right\}.
\end{eqnarray}
Then the fermionic part of the action is written as
\begin{eqnarray}
     S_F^{\bf S}
   &=&
       -\frac{\theta b_3v}{2}{\rm Tr}
           \left[
          - \hat{\chi}({\bf \hat{1}}-{\bf \hat{P}_S})
             \hat{\psi} ({\bf \hat{1}}-{\bf \hat{P}_S})
        +\hat{\chi}{\bf \hat{P}_S}\hat{\psi}{\bf \hat{P}_S}
          \right].
\end{eqnarray}
The partition function is given by
\begin{eqnarray}
    {\cal Z}_{\bf S}
 &=&
       \int
      {\cal D}\chi
       {\cal D}\psi
        {\cal D}\phi
           ~e^{-S}\nonumber\\
 &=&
           \int
            \prod_{\stackrel{\mbox {$\scriptscriptstyle i>j$}}
                  {\mbox{$\scriptstyle i,j\not\in S $}}}
             \frac{d\phi_{ij}}{\sqrt{4\pi\theta}}
              d\chi_{ij}d\psi_{ij}
               \prod_{\stackrel{\mbox {$\scriptscriptstyle i>j$}}
                     {\mbox{$\scriptstyle i,j\in S $}}}
                \frac{d\phi_{ij}}{\sqrt{4\pi\theta}}
                d\chi_{ij}d\psi_{ij}
            ~e^{-S_B^{\bf S}-S_F^{\bf S}}\nonumber\\
 &=&
              \left[\sqrt{(\frac{b_3v}{2})^2}
              \right]^{-(N-n)^2}
               \left[\sqrt{(\frac{ b_3v}{2})^2}
                \right]^{-n^2}
             \left[-\frac{b_3v}{2}
            \right]^{(N-n)^2}
             \left[\frac{b_3v}{2}
              \right]^{n^2},
\end{eqnarray}
and we get
\begin{eqnarray}
   {\cal Z}_{\bf S}
         &=&\left[{-\rm sgn}(b_3)~\right]^{(N-n)^2}
         \left[~{\rm sgn}(b_3)~\right]^{n^2}\nonumber\\
        &=&\left[-{\rm sgn}(b_3)~\right]^{N-n}
         \left[~{\rm sgn}(b_3)~\right]^{n}
\end{eqnarray}
where $n={\rm rank}S$. There are ${\scsc \frac{N!}{(N-n)!n!}}$ sets of the
GMS solution which
has the rank $n$. Then the total partition function is the sum over
${\cal Z}_{\bf S}$ with the weight ${\scsc \frac{N!}{(N-n)!n!}}$,
\begin{eqnarray}
     {\cal Z}_{Total}
    &=&\sum_{\bf S}~{\cal Z}_{\bf S} \nonumber \\
     &=&\lim_{N\rightarrow\infty}~
       ~\sum_{n=0}^{N}
 {}_NC_n  \left[-{\rm sgn}(b_3)~\right]^{N-n}
         \left[~{\rm sgn}(b_3)~\right]^{n}.
\end{eqnarray}
From the binomial theorem, we find that
\begin{eqnarray}
        {\cal Z}_{Total}=0.
\end{eqnarray}
In the $\phi^3$ potential, the partition function is zero.
%%%%%%%%%%%%%%%%%%%%%%%%%%%%%%%%%%%%%%%%%%%%%%%%%%%%%%%%%%%%%%%%%%%%%%%%%
\section{Large $\theta$ limit:$\phi^4$ potential}
\setcounter{equation}{0}
%%%%%%%%%%%%%%%%%%%%%%%%%%%%%%%%%%%%%%%%%%%%%%%%%%%%%%%%%%%%%%%%%%%%%%%%%
Here we consider another simple example which has $\phi^4$ potential,
\begin{eqnarray}
        V(\phi)=b_0 +b_1\phi
          +\frac{1}{2}b_2 \phi *\phi
           +\frac{1}{3!}b_3 \phi*\phi*\phi
            +\frac{1}{4!}b_4 \phi*\phi*\phi*\phi.
\end{eqnarray}
The field equation is always written as a factorization form
\begin{eqnarray}
         \frac{\delta V}{\delta\phi}
&=&\frac{1}{3!}b_4(\phi-\alpha)*(\phi-\beta)*(\phi-\gamma).
\end{eqnarray}
As is the same as the $\phi^3$, we translate the scalar field
\begin{eqnarray}
        \phi '\equiv \phi-\alpha.
\end{eqnarray}
Then the field equation is rewritten as
\begin{eqnarray}
     \frac{\delta V}{\delta\phi}
   &=&\frac{1}{3!}b_4 \phi '
      *\left(\phi ' -(\beta-\alpha)\right)
       *\left(\phi ' -(\gamma-\alpha)\right) \nonumber \\
      &=&\frac{1}{3!}b_4 \phi '
         *\left(\phi ' -v_1\right)
          *\left(\phi '-v_2\right),
\end{eqnarray}
where we define
\begin{eqnarray}
   v_1\equiv (\beta-\alpha),  v_2\equiv (\gamma-\alpha).
\end{eqnarray}
For simplicity we take $\alpha$, $\beta$ and $\gamma$ are real
numbers. We put the GMS solution $\phi '_{GMS}=\lambda_i {\bf P}_i$ into
the field equation, then
\begin{eqnarray}
      \frac{\delta V}{\delta\phi}
      =\frac{1}{3!}b_4\lambda_i
        \left(\lambda_i-v_1\right)
         \left(\lambda_i-v_2\right){\bf P}_i=0.
\end{eqnarray}
The solution is
\begin{eqnarray}
       \lambda_i =0, ~v_1, ~v_2.
\end{eqnarray}
The general GMS solution is a linear combination of projections with
coefficients $\lambda_i$,
\begin{eqnarray}
    \phi_{\scriptscriptstyle GMS}
 &=&
   v_1
       \sum_{i\in {\bf S_1}}{\bf P}_i
     +v_2
         \sum_{i\in {\bf S_2}}{\bf P}_i \\
 &=&
        v_1{\bf P_{S_1}}
         +v_2{\bf P_{S_2}}
\end{eqnarray}
The bosonic part of the action is given by
\begin{eqnarray}
 S_B^{\bf S}
 &=&
    -\frac{1}{4} {\rm Tr}
       \left[
             \left(\hat{\phi}_q\frac{\delta}{\delta \hat{\phi}}
              \right)
               \left.\frac{\delta V(\hat{\phi})}{\delta\hat{\phi}}
                \right|_{GMS}
           \left(\hat{\phi}_q\frac{\delta}{\delta \hat{\phi}}
            \right)
             \left.\frac{\delta V(\hat{\phi})}{\delta\hat{\phi}}
              \right|_{GMS}
        \right]
\end{eqnarray}
The linear terms of expansion of $V(\hat\phi)$ around the GMS soliton are
\begin{eqnarray}
    \lefteqn{\left(\hat\phi_q\frac{\delta}{\delta \hat\phi}
     \right)
      \left.\frac{\delta V(\hat\phi)}{\delta\hat\phi}
       \right|_{GMS}}\hspace{15mm} \nonumber\\
   &=&
    \left(\phi_q \frac{\delta}{\delta\hat\phi}
     \right)
      \left[\frac{b_4}{3!}\hat\phi (\hat\phi-v_1)(\hat\phi-v_2)
       \right]_{\hat\phi=\hat\phi_{GMS}}   \nonumber\\
   &=&
     \frac{b_4}{3!}
      \left\{
            \hat\phi_q
            (\hat\phi_{\scriptscriptstyle GMS}-v_1 )
             (\hat\phi_{\scriptscriptstyle GMS}-v_2)
               \right. \nonumber\\
   &&
      \hspace*{8mm} +\hat\phi_{\scriptscriptstyle GMS}
        \hat\phi_q
        (\hat\phi_{\scriptscriptstyle GMS}-v_2)
        +\left.
          \hat\phi_{\scriptscriptstyle GMS}
          (\hat\phi_{\scriptscriptstyle GMS}-v_2)
            \hat\phi_q
         \right\}\nonumber\\
   &=&
      \frac{b_4}{3!}
      \left\{
          (v_1 v_2)
             \hat\phi_q
              \left(\hat{\bf 1}
                    -{\bf \hat{P}_{S_1}}
                     -{\bf \hat{P}_{S_2}}
                 \right)\right.\nonumber \\
  &&
     \hspace*{8mm}  +\left(
           v_1 {\bf \hat{P}_{S_1}}
            v_2 {\bf \hat{P}_{S_2}} \right)
             \hat\phi_q  \nonumber\\
  &&
           \hspace*{13mm} \left[-v_2
               \left(\hat{\bf 1}
                    -{\bf \hat{P}_{S_1}}
                     -{\bf \hat{P}_{S_2}}
                 \right)
                 +\left(v_1 -v_2\right){\bf \hat{P}_{S_1}}
           \right] \nonumber\\
  &&
       \hspace*{8mm} +\left. v_2(v_2-v_1){\bf \hat{P}_{S_2}}
            \hat\phi_q\right\}.
\end{eqnarray}
The bosonic part of the action is
\begin{eqnarray}
 S_B^{\bf S}
     &=&
      -\frac{1}{4} \theta
           {\rm Tr}
           \left[~
              \left(b_4\zeta_0\right)^2
             \hat{\phi}_q
             \left(\hat{\bf 1}-{\bf \hat{P}_{S_1}}
                              -{\bf \hat{P}_{S_2}}\right)
               \hat{\phi}_q
             \left(\hat{\bf 1}-{\bf \hat{P}_{S_1}}
                              -{\bf \hat{P}_{S_2}}\right)\right.\nonumber\\
   &&
      \hspace*{15mm}+\left(b_4\zeta_1\right)^2
           {\bf \hat{P}_{S_1}}\hat{\phi}_q
           {\bf \hat{P}_{S_1}}\hat{\phi}_q
         +\left.\left(b_4\zeta_2\right)
               {\bf \hat{P}_{S_2}}\hat{\phi}_q
               {\bf \hat{P}_{S_2}}\hat{\phi}_q ~\right]
\end{eqnarray}
where we define
\begin{eqnarray}
  \zeta_0\equiv \frac{1}{3!}v_1 v_2,
 ~\zeta_1\equiv \frac{1}{3!}v_1 \left(v_1 -v_2\right),
  ~\zeta_2\equiv \frac{1}{3!}v_2  \left(v_2 -v_1\right).
\end{eqnarray}
The fermionic part of the Lagrangian is written as
\begin{eqnarray}
     S_F^{\bf S}
  &=&
        -i\theta{\rm Tr}
             \hat{\chi}
             \left(\hat{\psi}\frac{\delta}{\delta\hat{\phi}}
              \right)
         \left.\frac{\delta V(\hat{\phi})}{\delta\hat{\phi}}
          \right|_{GMS}.
\end{eqnarray}
The BRST operation is given by
\begin{eqnarray}
   \left(\hat\psi\frac{\delta}{\delta \hat\phi}
     \right)
      \left.\frac{\delta V(\hat\phi)}{\delta\hat\phi}
       \right|_{GMS}
    &=&
      \frac{b_4}{3!}
      \left\{
          (v_1 v_2)
             \hat\psi
             \left(\hat{\bf 1}
                    -{\bf \hat{P}_{S_1}}
                     -{\bf \hat{P}_{S_2}}
                 \right)\right.\nonumber \\
  &&
     \hspace*{8mm}  +\left(
           v_1 {\bf \hat{P}_{S_1}}
           +v_2 {\bf \hat{P}_{S_2}} \right)
             \hat\psi  \nonumber\\
  &&
           \hspace*{13mm} \times\left[-v_2
               \left(\hat{\bf 1}
                    -{\bf \hat{P}_{S_1}}
                     -{\bf \hat{P}_{S_2}}
                 \right)
                 +\left(v_1 -v_2\right){\bf \hat{P}_{S_1}}
           \right] \nonumber\\
  &&
       \hspace*{8mm} +\left. v_2(v_2-v_1){\bf \hat{P}_{S_2}}
            \hat\psi\right\}.
\end{eqnarray}
Then the fermionic part is
\begin{eqnarray}
     S_F^{\bf S}
  &=&-i\theta
       {\rm Tr}
           \left[
             b_4\zeta_0
                \hat{\chi}
                  \left(
                    {\bf \hat{1}}
                    -{\bf \hat{P}_{S_1}}
                     -{\bf \hat{P}_{S_2}}
                        \right)
                \hat{\psi}
                  \left(
                    {\bf \hat{1}}
                    -{\bf \hat{P}_{S_1}}
                     -{\bf \hat{P}_{S_2}}
                        \right)\right.\nonumber\\
        &&
         \hspace*{20mm}+b_4\zeta_1
               \hat{\chi}{\bf \hat{P}_{S_1}}
                \hat{\psi}{\bf \hat{P}_{S_1}} \nonumber\\
        &&
        \hspace*{20mm}  +b_4\left.\zeta_2
               \hat{\chi}{\bf \hat{P}_{S_2}}
                \hat{\psi}{\bf \hat{P}_{S_2}}\right].
\end{eqnarray}
The partition function is given
\begin{eqnarray}
   {\cal Z}_{\bf S}
 &=&
      \int {\cal D}
       \chi {\cal D}\psi{\cal D}\phi
         e^{-S}\\
 &=&
      \int \prod_{\bf S=S_0}^{S_2}
           [\prod_{\stackrel{\mbox {$\scriptscriptstyle m>n$}}
                   {\mbox{$\scriptstyle i,j\in {\bf S} $}}}
            \frac{d\phi_{ij}}{\sqrt{4\pi\theta}}
            d\chi_{ij}d\psi_{ij}
             ~e^{-S_B^{\bf S}-S_F^{\bf S}}
                    \nonumber\\
 &=&
       \times
        \left[\sqrt{(b_4\zeta_0)^2}
         \right]^{-(N-n_1-n_2)^2}
          \left[\sqrt{(b_4\zeta_1)^2}
           \right]^{-n_1^2}
            \left[\sqrt{(b_4\zeta_2)^2}
             \right]^{-n_2^2}
              \nonumber\\
  &&
        \times
         \left[b_4\zeta_0
          \right]^{(N-n_1-n_2)^2}
           \left[b_4\zeta_1
            \right]^{n_1^2}
             \left[b_4\zeta_2
              \right]^{n_2^2}.
\end{eqnarray}
Hence, we get
\begin{eqnarray}
    {\cal Z}_{\bf S}
   &=&
        \left[{\rm sgn }(b_4\zeta_0)
         \right]^{(N-n_1-n_2)^2}
          \left[{\rm sgn }(b_4\zeta_1)
           \right]^{n_1^2}
            \left[{\rm sgn }(b_4\zeta_2)
             \right]^{n_2^2}
\end{eqnarray}
where $n_1={\rm rank}S_1$, $n_2={\rm rank}S_2$. From the definition of
$\zeta_i$'s,
one of ${\rm sgn}(\zeta_i)$
is always negative, and the others are positive.
Then the partition function is always written as
\begin{eqnarray}
{\cal Z}_{\bf S}
   &=&
      \left[~{\rm sgn}(b_4)~\right]^{(N-n_1-n_2)^2}
             \left[-{\rm sgn}(b_4)~\right]^{n_1^2}
              \left[~{\rm sgn}(b_4)~\right]^{n_2^2}
            \\
&=&
    \left[~{\rm sgn}(b_4)~\right]^{N-n_1-n_2}
        \left[-{\rm sgn}(b_4)~\right]^{n_1}
         \left[~{\rm sgn}(b_4)~\right]^{n_2}
\end{eqnarray}
There are ${\scriptscriptstyle \frac{N!}{(N-n_1-n_2)!n_1! n_2!}}$ sets
of the GMS solution when the rank of ${\bf S}_1$ and ${\bf S}_2$ are
 $n_1$ and $n_2$. Then the total partition function is the sum over ${\bf
Z}_S$ with
the weight ${\scriptscriptstyle \frac{N!}{(N-n_1-n_2)!n_1! n_2!}}$ .
\begin{eqnarray}
     {\cal Z}_{Total}
    &=&
       \sum_{\bf S}~{\cal Z}_{\bf S}\nonumber \\
    &=&
       \lim_{N\rightarrow\infty}~
       ~\sum_{n_1, n_2=0}^{N}
        {\scriptstyle \frac{N!}{(N-n_1-n_2)!n_1! n_2!}}
          \left[
         ~{\rm sgn}(b_4)~\right]^{N-n_1-n_2}
           \left[
          -{\rm sgn}(b_4)
            ~\right]^{n_1}
              \left[
             ~{\rm sgn}(b_4)
               ~\right]^{n_2}\nonumber\\
   &=&
      \lim_{N\rightarrow\infty}
      ~\left(
       {\rm sgn}(b_4)-{\rm sgn}(b_4)
       +{\rm sgn}(b_4)\right)^N  \nonumber\\
        &=&
           \lim_{N\rightarrow\infty}~ \left({\rm sgn}(b_4)\right)^N.
\end{eqnarray}
In the $\phi^4$ potential, the partition function takes a non-vanishing
 value $\lim_{N\rightarrow\infty}~ \left({\rm sgn}(b_4)\right)^N$.
%%%%%%%%%%%%%%%%%%%%%%%%%%%%%%%%%%%%%%%%%%%%%%%%%%%%%%%%%%%%%%%%%
\section{Gaussian integral in the noncommutative space}
\setcounter{equation}{0}
%%%%%%%%%%%%%%%%%%%%%%%%%%%%%%%%%%%%%%%%%%%%%%%%%%%%%%%%%%%%%%%%%
In our calculations, the Gaussian functional integral appears, whose form is
\begin{equation}
 \int\mathcal{D}\phi\exp\left\{-\int d^2x\phi(x)
 \ast\left.V(\phi(x))\right|_{GMS}\ast\phi(x)\right\} ,
  \label{EQ:102}
\end{equation}
and we convert it into the number representation.
Generally, the difference of the operator ordering may yield
some difference of the result. Therefore we must show that our
prescription is correct.

Generally operators can be represented in the number representation as
\begin{equation}
 \hat{O}=\sum_{m,n}O_{mn}|m\rangle\langle n| ,
\end{equation}
where $\hat{O}$ is any operator and $|m\rangle\langle n|$ is the basis of
the number representation.
Any operator can be represented as the Weyl ordered operator,
so the basis $|m\rangle\langle n|$ can be written as
\begin{equation}
 |m\rangle\langle n|=\underbrace{
  \int\frac{d^2k}{(2\pi)^2}\tilde{f}_{mn}(k)
  e^{i(k_1\hat{x}^1+k_2\hat{x}^2)}}
  _{\mbox{\small{Weyl ordered operator}}} ,
\end{equation}
and its Weyl ordered symbol corresponding to the basis
$|m\rangle\langle n|$ is
\begin{equation}
 f_{mn}(x)
  =\int\frac{d^2k}{(2\pi)^2}\tilde{f}_{mn}(k)
  e^{i(k_1x^1+k_2x^2)} .
\end{equation}
Here, we consider the following integral,
\begin{eqnarray}
 & &\int d\hat{x}_1d\hat{x}_2|m\rangle\langle n|p\rangle\langle
q|\nonumber\\
 &=&\int d\hat{x}_1d\hat{x}_2
  \int\frac{d^2k}{(2\pi)^2}\frac{d^2k^{\prime}}{(2\pi)^2}
  \tilde{f}_{mn}(k)\tilde{f}_{pq}(k^{\prime})
  e^{i(k_1\hat{x}^1+k_2\hat{x}^2)}
  e^{i(k^{\prime}_1\hat{x}^1+k^{\prime}_2\hat{x}^2)}\nonumber\\
 &=&\int\frac{d^2k}{(2\pi)^2}\tilde{f}_{mn}(k)\tilde{f}_{pq}(-k) ,
\end{eqnarray}
where we set $[x^1,x^2]=1$ for simplicity and use
Baker-Campbell-Hausdorff formula.
On the other hand,
since $|m\rangle\langle n|p\rangle\langle q|=|m\rangle\delta_{np}\langle
q|$,
\begin{eqnarray}
 & &\int d\hat{x}_1d\hat{x}_2|m\rangle\delta_{np}\langle q|\nonumber\\
 &=&\int d\hat{x}_1d\hat{x}_2\int\frac{d^2k}{(2\pi)^2}
  \tilde{f}_{mq}(k)\delta_{np}e^{i(k_1\hat{x}^1+k_2\hat{x}^2)}\nonumber\\
 &=&\int d^2k\tilde{f}_{mq}(k)\delta^2(k)\delta_{np} .
\end{eqnarray}
Therefore we can derive a relation:
\begin{equation}
 \int\frac{d^2k}{(2\pi)^2}\tilde{f}_{mn}(k)\tilde{f}_{pq}(-k)
  =\int d^2k\tilde{f}_{mq}(k)\delta^2(k)\delta_{np} .\label{EQ:101}
\end{equation}
Using Eq.(\ref{EQ:101}), we get the following relation
\begin{equation}
 \int d^2x f_{mn}(x)\ast f_{pq}(x)=\int d^2x f_{mq}(x)\delta_{np} .
  \label{EQ:103}
\end{equation}
A similar calculation gives
\begin{eqnarray}
 \int d^2x f_{p_1q_1}(x)\ast f_{p_2q_2}(x)\ast\cdots\ast f_{p_nq_n}(x)
  =\int d^2x f_{p_1q_n}(x)\delta_{q_1p_2}\delta_{q_2p_3}
  \cdots\delta_{q_{n-1}p_n} .\label{EQ:104}
\end{eqnarray}

It is possible to expanded any function in terms of
$f_{mn}(x)$, so $\phi(x)$ and
$V(\phi)|_{GMS}$
in Eq.(\ref{EQ:102}) are expanded as
\begin{eqnarray}
        \phi(x)&=&\sum_{mn}\phi_{mn}f_{mn}(x),\\
 V(\phi)|_{GMS}&=&\sum_{l}V_{l}f_{ll}(x).
\end{eqnarray}
Therefore we perform the functional integral (\ref{EQ:102}) using
Eq.(\ref{EQ:103})(\ref{EQ:104}) as follows,
\begin{equation}
 \int\mathcal{D}\phi\exp
  \left\{-\int d^2x\phi_{mn}V_l\phi_{pq}(f_{mn}(x)\ast f_{ll}(x)\ast f_{pq})
 \right\}
 =\left[\det|V_l|^{\frac{1}{2}}\right]^{-1} .\label{EQ:105}
\end{equation}
On the other hand, a calculation in the number representation gives
\begin{eqnarray}
 & &\int\mathcal{D}\hat{\phi}\exp
  \left\{-{\rm Tr}\phi_{mn}V_l\phi_{pq}
   (|m\rangle\langle n|l\rangle\langle l|p\rangle\langle
q|)\right\}\nonumber\\
 &=&\int\mathcal{D}\phi\exp(-\phi_{ml}V_l\phi_{lm})\nonumber\\
 &=&\left[\det|V_l|^{\frac{1}{2}}\right]^{-1} ,
\end{eqnarray}
which is the same result as Eq.(\ref{EQ:105}) without normalizing
constant which is canceled out. Therefore we can perform integration in both
representation, and the number operator is used in our calculations.
%%%%%%%%%%%%%%%%%%%%%%%%%%%%%%%%%%%%%%%%%%%%%%%%%%%%%%%%%%%%%%%%%%
\section{$\$_n=0$}
\setcounter{equation}{0}
%%%%%%%%%%%%%%%%%%%%%%%%%%%%%%%%%%%%%%%%%%%%%%%%%%%%%%%%%%%%%%%%%%
We prove a theorem here.

\begin{figure}[t]
 \begin{center}
  \psfragscanon
  \psfrag{y}[][][2.0]{$\$_n$}
  \psfrag{x0}[][][1.5]{$v_0$}
  \psfrag{x1}[][][1.5]{$v_1$}
  \psfrag{x2}[][][1.5]{$v_2$}
  \psfrag{x3}[][][1.5]{$v_{n-2}$}
  \psfrag{x4}[][][1.5]{$v_{n-1}$}
  \psfrag{x5}[][][1.5]{$v_n$}
  \scalebox{.5}{\includegraphics{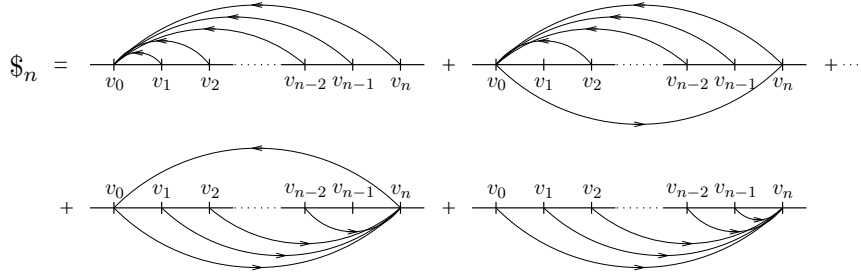}}
 \end{center}
 \caption{Graphical representation of $\$_n$.
 The arrow $v_i\leftarrow v_j$ means $(v_j-v_i)$, and all arrows in one
 figure should be multiplied.}
 \label{FIG:001}
\end{figure}

The theorem which we will prove here is
\begin{equation}
 \sum_{i=B}^A\prod_{\alpha =0}^{i-1}(v_A-v_{\alpha})
  \prod_{\beta =i+1}^{r}(v_B-v_{\beta})=0 ,
\end{equation}
where integers $A,B$ satisfy $0\leqslant B<A\leqslant r$, and $v_i$ is
any real number. We defined
the following notation formally for convenience,
\begin{equation}
 \prod_{\alpha =0}^{-1}(v_A-v_{\alpha})
  =\prod_{\alpha =r+1}^{r}(v_A-v_{\alpha})=1 .
\end{equation}
Its proof is as follows.\\
\underline{Proof}
\begin{eqnarray}
 & &\sum_{i=B}^A\prod_{\alpha =0}^{i-1}(v_A-v_{\alpha})
  \prod_{\beta =i+1}^{r}(v_B-v_{\beta})\nonumber\\
 &=&\sum_{i=B}^A(v_A-v_0)(v_A-v_1)\cdots(v_A-v_{i-1}){}^{\vee}
  (v_B-v_{i+1})(v_B-v_{i+2})(v_B-v_m)\nonumber\\
 &=&\underbrace{(v_A-v_0)(v_A-v_1)\cdots(v_A-v_{B-1})}_{B}
  \underbrace{(v_B-v_{B+1})(v_B-v_{B+2})\cdots(v_B-v_m)}_{r-B}\nonumber\\
 & &+\underbrace{(v_A-v_0)(v_A-v_1)\cdots(v_A-v_{B})}_{B+1}
  \underbrace{(v_B-v_{B+2})(v_B-v_{B+3})\cdots(v_B-v_m)}_{r-B-1}\nonumber\\
 & &+\cdots\nonumber\\
 & &+\underbrace{(v_A-v_0)(v_A-v_1)\cdots(v_A-v_{A-1})}_{A}
  \underbrace{(v_B-v_{A+1})(v_B-v_{A+2})\cdots(v_B-v_m)}_{r-A}\nonumber
\end{eqnarray}
\begin{eqnarray}
 &=&(v_A-v_0)(v_A-v_1)\cdots(v_A-v_{B-1})\cdot
     (v_B-v_{A+1})(v_B-v_{A+2})\cdots(v_B-v_m)\nonumber\\
 & &\times\{1\cdot(v_B-v_{B+1})(v_B-v_{B+2})\cdots(v_B-v_A)\nonumber\\
 &
&\;\;\;\;+(v_A-v_B)\cdot(v_B-v_{B+2})(v_B-v_{B+3})\cdots(v_B-v_A)\nonumber\\
 & &\;\;\;\;+\cdots\nonumber\\
 & &\;\;\;\;+(v_A-v_B)(v_A-v_{B+1})\cdots(v_A-v_{A-1})\cdot1\}
.\label{EQ:201}
\end{eqnarray}
To show that $\{\cdots\}$ in Eq.(\ref{EQ:201}) equals to zero,
we define $\$_n$ as
(Fig.\ref{FIG:001})
\begin{eqnarray}
 \$_n&=&\;\;\;(v_0-v_1)(v_0-v_2)(v_0-v_3)\cdots(v_0-v_n)\nonumber\\
 & &+(v_n-v_0)\cdot(v_0-v_2)(v_0-v_3)\cdots(v_0-v_n)\nonumber\\
 & &+(v_n-v_0)(v_n-v_1)\cdot(v_0-v_3)\cdots(v_0-v_n)\nonumber\\
 & &+\cdots\nonumber\\
 & &+(v_n-v_0)\cdots(v_n-v_{n-3})\cdot(v_0-v_{n-1})(v_0-v_n)\nonumber\\
 & &+(v_n-v_0)\cdots(v_n-v_{n-3})(v_n-v_{n-2})\cdot(v_0-v_n)\nonumber\\
 & &+(v_n-v_0)\cdots(v_n-v_{n-3})(v_n-v_{n-2})(v_n-v_{n-1}) .
\end{eqnarray}
\begin{figure}[t]
 \begin{center}
  \psfragscanon
  \psfrag{y}[-1][][2.0]{$\$_{n+1}$}
  \psfrag{x0}[][][1.5]{$v_0$}
  \psfrag{x1}[][][1.5]{$v_1$}
  \psfrag{x2}[][][1.5]{$v_2$}
  \psfrag{x3}[][][1.5]{$v_{n-1}$}
  \psfrag{x4}[][][1.5]{$v_{n+1}$}
  \psfrag{x5}[][][1.5]{$v_n$}
  \scalebox{.5}{\includegraphics{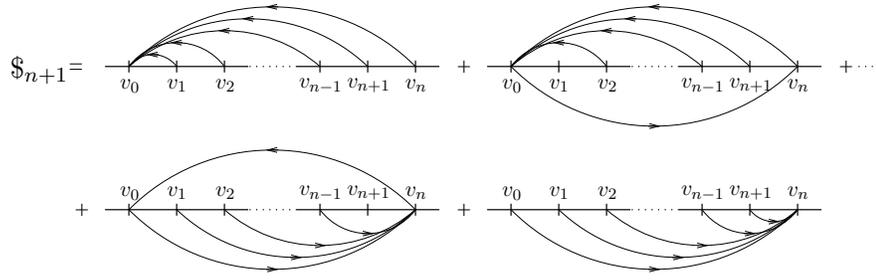}}
 \end{center}
 \caption{Graphical representation of $\$_{n+1}$.
 We put $v_{n+1}$ between $v_{n-1}$ and $v_n$ in the proof.}
 \label{FIG:002}
\end{figure}
We prove $\$_n=0$ using mathematical induction as follows.
First, for $n=1$,
\begin{equation}
 \$_1=(v_0-v_1)+(v_1-v_0)=0 .
\end{equation}
Next, we suppose $\$_n=0$. And we consider $\$_{n+1}$ as (Fig.\ref{FIG:002})
\begin{eqnarray}
 \$_{n+1}&=&\;\;\;(v_0-v_1)(v_0-v_2)(v_0-v_3)\cdots
  (v_0-v_{n-1})(v_0-v_{n+1})(v_0-v_n)\nonumber\\
 & &+(v_n-v_0)\cdot(v_0-v_2)(v_0-v_3)\cdots
  (v_0-v_{n-1})(v_0-v_{n+1})(v_0-v_n)\nonumber\\
 & &+(v_n-v_0)(v_n-v_1)\cdot(v_0-v_3)\cdots
  (v_0-v_{n-1})(v_0-v_{n+1})(v_0-v_n)\nonumber\\
 & &+\cdots\nonumber\\
 & &+(v_n-v_0)\cdots(v_n-v_{n-2})\cdot(v_0-v_{n+1})(v_0-v_n)\nonumber\\
 & &+(v_n-v_0)\cdots(v_n-v_{n-2})(v_n-v_{n-1})\cdot(v_0-v_n)\nonumber\\
 & &+(v_n-v_0)\cdots(v_n-v_{n-2})(v_n-v_{n-1})(v_n-v_{n+1})\nonumber\\
 &=&\{\$_n-(v_n-v_0)(v_n-v_1)\cdots(v_n-v_{n-1})\}(v_0-v_{n+1})
  \nonumber\\
 & &+(v_n-v_0)(v_n-v_1)\cdots(v_n-v_{n-1})
  \{(v_0-v_n)+(v_n-v_{n+1})\}\nonumber\\
 &=&(v_n-v_0)(v_n-v_1)\cdots(v_n-v_{n-1})
  \{-(v_0-v_{n+1})+(v_0-v_n)+(v_n-v_{n+1})\}\nonumber\\
 &=&0 .
\end{eqnarray}
Therefore $\$_n=0$ is valid for any $n$.
Using it, $\{\cdots\}$ in Eq.(\ref{EQ:201}) equals to zero,
so the proof is completed.
%%%%%%%%%%%%%%%%%%%%%%%%%%%%%%%%%%%%%%%%%%%%%%%%%%%%%%%%%%%%%%%%%%%%%%%%%
%%%%%%%%%%%%%%%%%%%%      bibliography   %%%%%%%%%%%%%%%%%%%%%%%%%%%%%%%%
%%%%%%%%%%%%%%%%%%%%%%%%%%%%%%%%%%%%%%%%%%%%%%%%%%%%%%%%%%%%%%%%%%%%%%%%%

%%%%%%%%%%%%%%%%%%%%%%%%%%%%%%%%%%%%%%%%%%%%%%%%%%%%%%%%%%%%%%%%%%%%%%%%%
%%%%%%%%%%%%%%%%%%%%%%%%%%%%%%%%%%%%%%%%%%%%%%%%%%%%%%%%%%%%%%%%%%%%%%%%%
\end{document}